\documentclass[twocolumn,secnumarabic,assume, nobibnotes, aps, prd, floatfix]{revtex4-2}
\usepackage{graphics, graphicx}
\usepackage{bm}
\usepackage{subscript}
\usepackage{mathtools,amsmath}
\usepackage{adjustbox}
\usepackage{array, booktabs,tabularx}
\usepackage{siunitx}
\usepackage{upquote}
\usepackage{hhline}
\usepackage{stackengine}
\usepackage{xcolor}
\usepackage{setspace} 
\usepackage{hyperref}
\usepackage{microtype}
\usepackage[capitalize]{cleveref}
\usepackage{xr}
\usepackage{float}
\usepackage{xspace}
\usepackage{graphicx}
\usepackage{subcaption}
\usepackage[customcolors]{hf-tikz}
\usepackage{gensymb}
\usepackage[version=4]{mhchem}

\usepackage{tabularx}

\newcommand{\MnBi}{Mn\textsubscript{Bi}\xspace}
\newcommand{\BiMn}{Bi\textsubscript{Mn}\xspace}

\newcommand{\MBT}{MnBi\textsubscript{2}Te\textsubscript{4}\xspace}

\newcommand{\sfig}[1]{Fig.~S#1}

\newcommand{\stabs}[2]{Tables~S#1 and~S#2}

\newcommand{\sfigsRange}[2]{Figs.~S#1 to~S#2}

\begin{document}

\begin{abstract}

The magnetic material \MBT (MBT) has garnered significant attention due to its unique combination of long-range antiferromagnetism and nontrivial topological electronic properties.
However, experimental measurements report inconsistent magnetizations,
which could be attributed to a variety of 
intrinsic defects.
To date,
a comprehensive investigation of 
defect-engineered MBT systems has not yet been established.
Employing state-of-the-art $ab~initio$ techniques, 
this work systematically investigates the influence of various experimentally reported defects on the magnetic properties of bulk and monolayer MBT at different concentrations.
Mn-vacancy and Mn-rich defects are found to enhance the ferromagnetism of bulk MBT. 
The investigation of Mn-rich and intermixing defects in the monolayer reveals that subtle structural and electronic modifications can alter the magnetic coupling.
Projection onto a Heisenberg Hamiltonian demonstrates that defects modify exchange interactions, thereby giving rise to distinct magnetic ground states.
This work sheds light on magnetic coupling mechanisms and
provides guidelines for the experimental control of magnetism in \MBT.

\end{abstract}

\title{\textit{Ab initio} study of magnetism in pristine and defective \texorpdfstring{$\mathbf{MnBi_2Te_4}$}{MnBi2Te4}}%

\author{Ana Beatriz Pedro Fontes}
\email{ana.pedrofontes@uclouvain.be}
\affiliation{Institute of Condensed Matter and Nanosciences (IMCN), Université catholique de Louvain (UCLouvain), B-1348, Louvain-la-Neuve, Belgium}

\author{Jiaqi Zhou}
% \email{jiaqi.zhou@uclouvain.be}
\affiliation{Institute of Condensed Matter and Nanosciences (IMCN), Université catholique de Louvain (UCLouvain), B-1348, Louvain-la-Neuve, Belgium}

\author{Simon M.-M. Dubois}
% \email{simon.dubois@uclouvain.be}
\affiliation{Institute of Condensed Matter and Nanosciences (IMCN), Université catholique de 
Louvain (UCLouvain), B-1348, Louvain-la-Neuve, Belgium}

\author{Jean-Christophe Charlier}
% \email{jean-christophe.charlier@uclouvain.be}
\affiliation{Institute of Condensed Matter and Nanosciences (IMCN), Université catholique de Louvain (UCLouvain), B-1348, Louvain-la-Neuve, Belgium}

\date{\today}%
\maketitle

\section{Introduction}

Magnetization control is essential
for the development of next-generation spintronic devices
with low power consumption and high-speed switching properties~\cite{fert2024electrical}.
Van der Waals (vdW) layered materials provide an attractive platform for this purpose owing to their exfoliability, atomic thinness, and tunable electronic properties, which can be further engineered through gate voltage and heterostructure design~\cite{Shen2022Dec, Peng2019Aug,zhou2021controllable,Zatko2022Sep}.
Among them, \MBT (MBT) stands out as the first experimentally realized vdW layered material that intrinsically combines long-range magnetic order with a topological insulating state~\cite{doi:10.1126/sciadv.aaw5685, Otrokov2019Mar, Zhang2019May}.
By tuning the layer thickness, externally applied magnetic field, strain, and chemical composition, MBT can exhibit exotic phenomena such as the quantum anomalous Hall effect, axion insulator states, and Weyl semimetallic behavior~\cite{Deng2020Jan, Li2024, Liu2020May}. These phenomena give rise to highly desirable properties, including quantized transverse conductivity, magnetoelectric effects, and dissipationless edge states, establishing MBT as a key material for both fundamental studies in topological quantum matter and prospective applications in spintronics and quantum information technologies~\cite{Li2024, Tokura2019Feb}. 

Although theoretical models predict robust topological phases, experiments on bulk or thin-film MBT deviate from the expected behaviors~\cite{PhysRevLett.132.066604}.
Angle-resolved photoemission spectroscopy has revealed a surface Dirac gap consistent with topological magnetism~\cite{PhysRevResearch.1.012011}. 
In contrast, other works using the same experimental technique detected significantly reduced or even gapless surface states~\cite{Garnica2022,PhysRevX.9.041038}. 
These conflicting results were attributed to the high sensitivity of the magnetic and electronic configurations of MBT to lattice imperfections and chemical inhomogeneities~\cite{hu2024recent}.
Many mechanisms
have been proposed
for the interpretation, 
including near-surface magnetization dead layers, deviations from the bulk magnetic order, surface collapse of the vdW gap, and intrinsic defects~\cite{vyazovskaya2025intrinsic}. 
Among the various intrinsic defects, 
antisite disorders 
involving Mn atoms substituting Bi sites (\MnBi)
and
Bi atoms occupying Mn sites (\BiMn) are particularly significant. 
These defects 
often form an intermixing disorder (\MnBi-\BiMn)
and
are suggested to be the most plausible cause
of the reduction of
the Dirac energy gap~\cite{Garnica2022,yan2022perspective, Li2024Jun}. 

Structural characterization confirms the high prevalence of these intrinsic defects. Scanning transmission electron microscopy and X-ray diffraction have revealed substantial defect concentrations in MBT, including \MnBi antisites at 2.5–6\%, \BiMn antisites reaching up to 15\%, and 
Mn vacancies around 4.9\%~\cite{Garnica2022,yan2022perspective, Li2024Jun,PhysRevResearch.1.012011,Zeugner2019}.
These defects distort the local lattice environment and influence both the intra- and inter-layer magnetic couplings. 
In particular, \MnBi antisites have been shown to induce ferrimagnetism in MBT by introducing antiferromagnetic coupling to the pristine Mn layer, thereby converting the intralayer exchange from ferromagnetic to ferrimagnetic~\cite{Lai2021, advs.202402753}.
More generally, structural distortions induced by intrinsic defects reshape the Mn–Te–Mn bonding geometry, modifying the superexchange pathways that determine the magnetic order~\cite{doi:10.1126/sciadv.aaw5685, Li2019Oct}. 
The above discussion underscores the strong influence of defects on the magnetic ground state of MBT.
However, prior studies remain limited in both defect-type diversity and concentration dependence~\cite{Garnica2022,doi:10.1021/acs.nanolett.3c00956,advs.202402753,PhysRevMaterials.4.121202}. A systematic analysis covering a broader class of intrinsic defects is still lacking,
and the defect-induced modulation of exchange interactions has yet to be elucidated.
In addition, the connection between local structural distortions and the emergence of long-range magnetic order remains unclear.

In this work, we present a systematic investigation of the defect-induced influence on the magnetism of MBT using first-principles calculations. 
A set of experimentally observed intrinsic defects,
including \BiMn, \MnBi, Mn-vacancy, and intermixing,
has been considered for both bulk and monolayer MBT
at various concentrations.
By comparing the total energies of the ferromagnetic and antiferromagnetic configurations,
the magnetic ground states can be identified in these defective systems. 
Additionally, to elucidate the underlying microscopic mechanism, we mapped the exchange parameters of two representative monolayer systems.
Our analysis highlights the interplay between crystal lattice distortions, modified exchange pathways and electronic structure in determining the overall magnetic behavior of MBT.

\section{Methods}

Calculations were performed within the framework of density functional theory (DFT) using the projector augmented wave (PAW) method implemented in the Vienna \textit{ab initio} simulation package (VASP)~\cite{Kresse1999Jan,Kresse1996}.
The Perdew-Burke-Ernzerhof~\cite{Perdew1996Oct} parameterization of the generalized gradient approximation (GGA-PBE) was used to describe the exchange-correlation interactions.
A kinetic energy cutoff of 350~eV for the plane waves was used in the calculations after thorough convergence testing, as detailed in \sfig{1} in the Supplemental Material (SM)~\cite{supplement2025}.
The Mn 3$d$ states were treated using the GGA+$U$ approach with a $U$ value of 4~eV in the Liechtenstein scheme~\cite{Liechtenstein1995Aug}, a value consistent with previous literature~\cite{doi:10.1126/sciadv.aaw5685, Li2020, Li2021Apr} and validated in \stabs{1}{2} in the SM~\cite{supplement2025}.
The DFT-D3 method~\cite{Grimme2010Apr} was employed to model the vdW interlayer interactions.
Structural optimizations were performed within a collinear spin-polarized framework, neglecting spin-orbit coupling (SOC). 
For the pristine bulk MBT, variable-cell structural relaxations yielded lattice constants of $a = b = 4.36$~{\AA} and $c = 27.59$~{\AA}, with angles $\alpha = \beta = 80.91^\circ$ and $\gamma = 60^\circ$.
For the monolayer, the optimized structural parameters were $a = b = 4.34$~{\AA} ($\alpha = \beta = 90^\circ$, $\gamma = 120^\circ$). A vacuum layer of 15~{\AA} was set along the $c$-axis to mitigate spurious image interactions.
To investigate the defective structures, supercells were constructed from these optimized pristine lattices. The bulk systems were modeled using $2 \times 2 \times 1$ and $3 \times 3 \times 1$ supercells, while $2 \times 2 \times 1$, $3 \times 3 \times 1$, and $4 \times 4 \times 1$ supercells were utilized for the monolayer.
In all defective systems, the lattice vectors were kept fixed to the pristine values, while the internal atomic positions were relaxed until the forces on each atom were below $0.01$~eV/{\AA}. 
Self-consistent calculations were performed with a total energy convergence of $10^{-6}$~eV, including SOC within a noncollinear magnetic framework.
In these calculations, the Mn magnetic moments were initialized along the $\mathrm{[001]}$ direction, remaining nearly aligned along the $c$-axis with a magnitude of approximately $4.5~\mu_{\rm B}$ per Mn ion.
For bulk systems, $k$-mesh samplings of $6\times6\times6$ and $4\times4\times2$ were adopted for $2\times2\times1$ and $3\times3\times1$ supercells, respectively. In the monolayer case, a uniform $k$-point density of 0.14~\AA$^{-1}$ was used in the $ab$-plane.

Two representative configurations, namely the Mn-rich and intermixing MBT monolayers at a defect concentration of 11.1\%, were chosen to elucidate the defect-induced modulation of exchange interactions.
To build the tight-binding model for extracting the exchange parameters,
noncollinear DFT calculations were performed using the SIESTA~\cite{Soler2002Mar} code (version 5.4.1) based on the VASP-optimized geometries. 
We used a double-zeta-polarized basis set automatically generated by SIESTA and norm-conserving, fully relativistic pseudopotentials from the PseudoDojo database~\cite{VANSETTEN201839}.
The exchange-correlation interactions were treated using the PBE~\cite{Perdew1996Oct} parameterization, and the Mn 3$d$ states were addressed with a $U$ value of 4 eV, using an identical method to VASP.
The consistency between the VASP and SIESTA frameworks was verified by comparing the total energy differences of the investigated systems.
To ensure the convergence of exchange parameters, we utilized a mesh-cutoff of 800~Ry and $k$-mesh sampling of $0.05$~\AA$^{-1}$ in the SIESTA calculations.
The exchange parameters were subsequently extracted using the Liechtenstein-Katsnelson-Antropov-Gubanov (LKAG)~\cite{LIECHTENSTEIN198765} formalism, as implemented in the TB2J~\cite{HE2021107938} code (version 0.9.12.26),
with $k$-mesh sampling identical to that of the SIESTA calculations.

\section{Results and discussions}
In the crystal structure, MBT is composed of stacked septuple layers (SLs), each consisting of a sequence of seven atomic planes (Te–Bi–Te–Mn–Te–Bi–Te) arranged along the crystallographic $c$-axis and held together by  weak interlayer vdW forces.
Within each SL, the Mn atoms form a ferromagnetic (FM) sublattice with an out-of-plane spin orientation.
The coupling between adjacent SLs is weakly antiferromagnetic (AFM), resulting in an overall A-type AFM structure~\cite{Yan2019Jun, Otrokov2019Dec}. 

\begin{figure}[t!]
    \centering
    \includegraphics[width=3.4in]{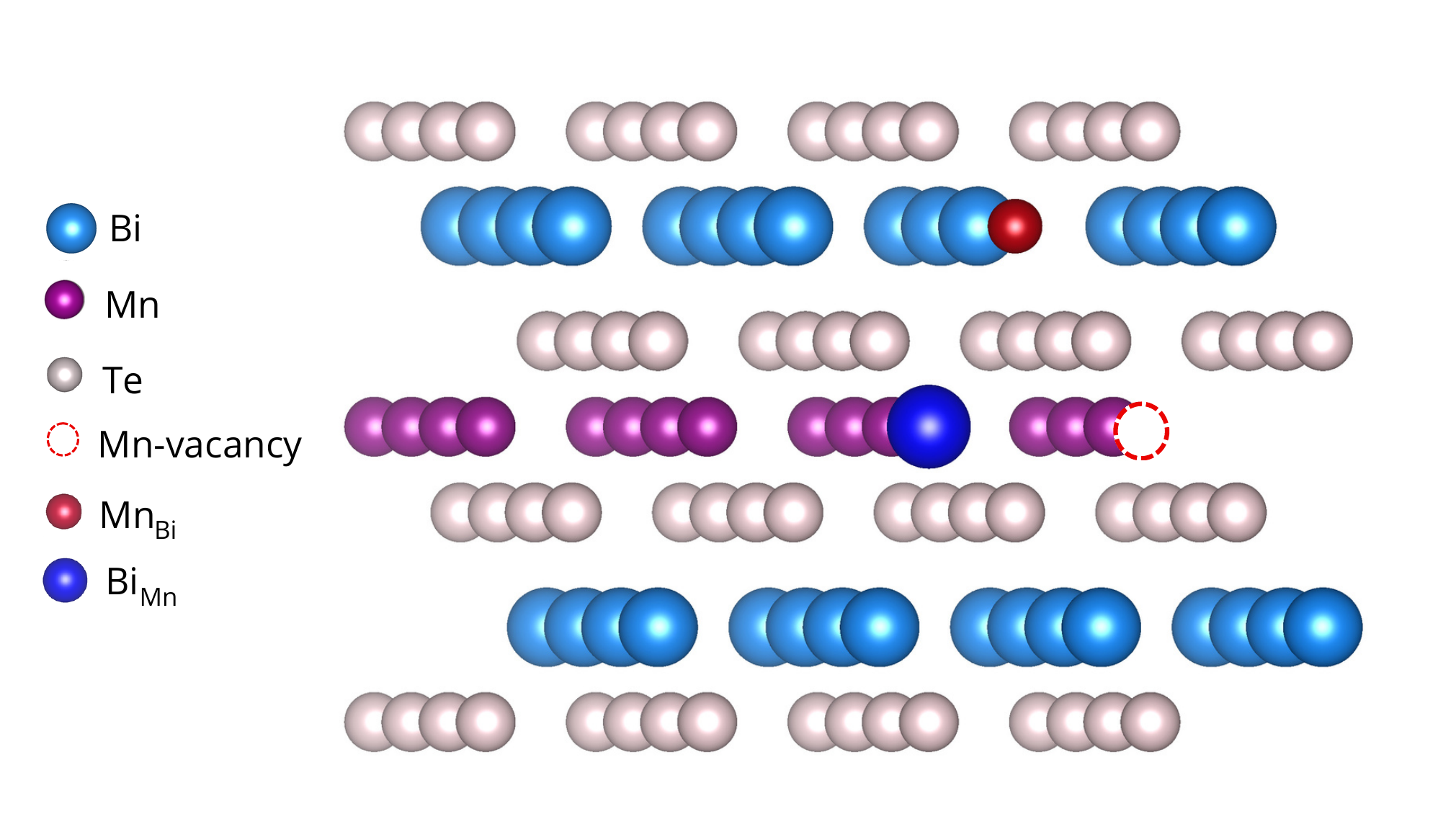}
    \captionsetup{justification=justified}
    \caption{\MBT septuple layer with various defects: Mn-vacancy (red dashed circle), \MnBi antisite (Mn substituting Bi - red small sphere), \BiMn antisite (Bi substituting Mn - dark blue large sphere), and the intermixing of Bi and Mn atoms
    where \MnBi and \BiMn simultaneously occur.
    The Bi, Mn, and Te atoms of MBT are illustrated using light blue, purple, and grey spheres, respectively.
    }
    \label{fig:defects}
\end{figure}

Based on the intrinsic defects observed in the experimental MBT samples~\cite{Garnica2022,hou2020te}, a set of defective models were constructed:
Mn-vacancy, Mn-rich, Bi-rich, and intermixing. 
\Cref{fig:defects} illustrates all investigated defects within a single septuple layer.
In the Mn-vacancy model, one Mn atom is removed from the central Mn layer.
The Mn-rich model,
denoted by the \MnBi defect, corresponds to a Mn atom occupying the Bi site, while the Bi-rich model involves a Bi atom 
occupying a Mn site, abbreviated as \BiMn.
In the intermixing model, the Bi and Mn sites are swapped.

\subsection{Interlayer and intralayer exchange coupling in \texorpdfstring{$\mathbf{MnBi_2Te_4}$}{MnBi2Te4} bulk}

DFT calculations were performed to determine the magnetic ground state of the bulk MBT. 
A canted cell containing two SLs was employed to investigate both the FM and AFM configurations of pristine MBT. 
The defect concentration is defined as the number of defects per SL per unit cell. 
Specifically, $2\times2\times1$ and $3\times3\times1$
supercells with one defect were employed,
corresponding to defect
concentrations of 12.5\% and 5.5\%, respectively. 
For all models, the total energies of the AFM and FM configurations were calculated.
For the Mn-rich and intermixing defect models, both antiferrimagnetic (AFiM) and ferrimagnetic (FiM) configurations were examined due to the presence of additional Mn atoms,
namely, \MnBi antisite defects in the Bi layer. The atomic models of the magnetic configurations
studied are explicitly provided in \cref{tab:configurations_bulk} of the Appendix section.

\Cref{fig:combined} shows the energy difference (\(\Delta E_{\mathrm{Bulk}}\)) between the AFM and FM states for all models analyzed, defined as
\begin{equation}
  \label{eq:deltaE}
  \Delta E_{\mathrm{Bulk}} = E_{\mathrm{FM}} - E_{\mathrm{AFM}},
\end{equation}
where $E_{\mathrm{AFM}}$ and $E_{\mathrm{FM}}$ are the total energies of the AFM and FM phases, respectively. The energy difference was calculated using the lowest-energy relaxed structure.
For a consistent comparison across different supercell sizes, these energies have been normalized per SL per unit cell.
A positive (negative) $\Delta E_{\mathrm{Bulk}}$ indicates that the AFM (FM) ordering is energetically favored. 
Specifically, in the Mn-rich and intermixing defect models, both A-type AFM and AFiM orderings are possible. In such cases, the AFM energy was taken as the lowest value as
\begin{equation}
  \label{eq:EAFM}
  \begin{aligned}
  E_{\mathrm{AFM}} = \mathrm{min}(E_{\mathrm{AFM}}, E_{\mathrm{AFiM}}).
  \end{aligned}
\end{equation}
Although FiM states were considered, they never yielded the lowest energy and are therefore omitted from the discussion. 
The $\Delta E_{\mathrm{Bulk}}$ of pristine MBT is presented as a reference
with a ground state of AFM.

\begin{figure}[t!]
    \centering
    \includegraphics[width=0.45\textwidth]{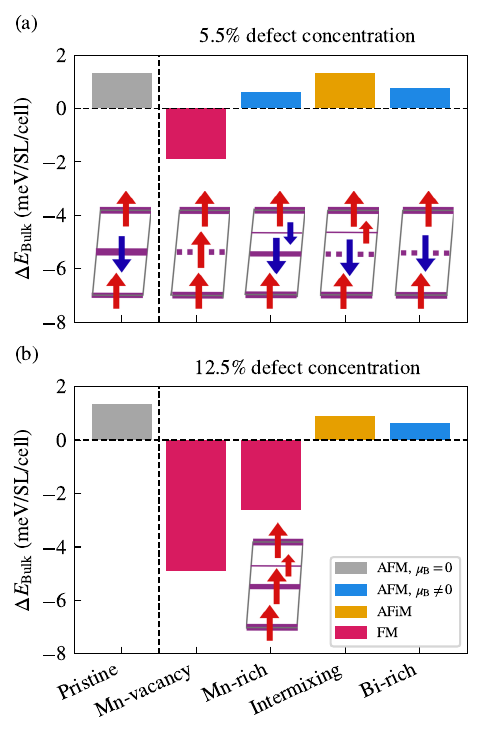}
    \caption{Interlayer coupling for pristine and defective supercell models containing (a) 5.5\% and (b) 12.5\% defect concentrations. 
    The schematic models depict the magnetic ground state of the defect-containing septuple layer for each model, 
    with the defect types given on the horizontal axis.
    Red (blue) upward (downward) arrows represent magnetic moments oriented upward (downward) within a given layer. The solid purple lines indicate Mn layers without defects, the dashed line indicate Mn layers missing one magnetic atom, and the slim purple line in the Mn-rich and intermixing models signifies the presence of an additional Mn atom.}
    \label{fig:combined}
    \begin{minipage}[t]{0.49\textwidth}
      \phantomsubcaption
      \label{fig:FMcoupling55}
  \end{minipage}
  \begin{minipage}[t]{0.49\textwidth}
      \phantomsubcaption
      \label{fig:FMcoupling125}
  \end{minipage}
\end{figure}

The computed layer-resolved magnetic moments are schematically depicted in \cref{fig:combined}.
Full magnetic moment compensation ($\mu_\mathrm{B}=0$) is achieved only in the pristine and intermixing models where the stoichiometry is preserved. 
In other defective systems, the interlayer coupling remains AFM, but the total magnetic moment becomes finite due to the broken stoichiometry.
This nonzero magnetic moment stems from incomplete cancellation in the Mn layers, caused by either a Mn vacancy (dashed line) or an additional Mn atom (slim line in Mn-rich and intermixing models).

With the introduction of a Mn-vacancy concentration of 5.5\%,
FM interlayer coupling is induced
between the two main Mn layers. 
Increasing the concentration of Mn vacancies to 12.5\%
further enhances this ferromagnetism, 
demonstrating the efficacy of Mn vacancies in promoting FM order.
Including one Mn atom in the adjacent Bi layer
makes the latter a defective layer with magnetization,
thus 
the Mn-rich and 
the \MnBi-\BiMn intermixing models
allow the exchange interaction between the main and defective layers
within one SL,
i.e., intralayer magnetic coupling.
In the Mn-rich model with 5.5\% defect concentration, 
the Mn at the Bi site 
aligns ferromagnetically with the main Mn layer 
inside the SL, 
leading to intralayer FM coupling. 
This reduces the AFM strength 
relative to pristine MBT. 
When the Mn-rich concentration increases to 12.5\%, 
the AFM phase is fully converted to the FM phase.  
In contrast, the ground state of \MnBi-\BiMn intermixing 
exhibits AFM intralayer and interlayer coupling
at both concentrations,
denoted as AFiM.
Besides, 
the Bi-rich models weaken the AFM strength at both concentrations.
These results indicate that incorporating Mn vacancies and Mn atoms 
into the \MBT bulk is an effective strategy to enhance FM coupling.
Relaxation of the internal atomic coordinates for all defective configurations largely preserves magnetic couplings,
as given in \sfig{2} of the SM~\cite{supplement2025}.
 
Our theoretical findings are in agreement with experimental studies.
Among the investigated defects, intermixing is the most prevalent and experimentally accessible defect~\cite{Garnica2022,doi:10.1021/acs.nanolett.3c00956,advs.202402753}.
It has been reported that the magnetic moments of Mn sites in the Mn layer and Mn antisites in the Bi layer (\MnBi) have an opposite alignment, leading to FiM intralayer coupling~\cite{advs.202402753}.
These findings, along with observations under a high-field magnetization~\cite{Lai2021}, verify the AFiM ground state reported by our calculations for the intermixing-defect systems where the FiM layers are coupled antiferromagnetically.
Moreover, another chemically analogous compound, $\mathrm{MnSb_2Te_4}$ with a higher concentration of Mn, has shown an FM coupling of Mn antisites with the Mn layer~\cite{wimmer2021mn}.
Since $\mathrm{MnSb_2Te_4}$ and $\mathrm{MnBi_2Te_4}$
belong to the same compound family, this observation 
provides a 
comparable framework
for demonstrating the enhancement of FM induced by the Mn-rich condition. 

\subsection{Intralayer exchange coupling in \texorpdfstring{$\mathbf{MnBi_2Te_4}$}{MnBi2Te4} monolayer}
The intricate three-dimensional magnetic order in bulk MBT arises from both intralayer exchange and vdW interlayer interactions. The monolayer serves as a simplified model for understanding the intraplanar mechanisms of the bulk material, as it removes the interlayer interactions. 
By studying a single septuple layer, the effects of defects on intralayer interactions are investigated without interference from adjacent layers.
Recent studies have shown that single-ion anisotropy and intralayer exchanges remain consistent across the \MBT family~\cite{Li2025Feb}, supporting our focus on the monolayer.

To evaluate the concentration dependence of the magnetic properties, we introduced a single defect into the supercells previously described, corresponding to defect concentrations of $25.0 \%$, $11.1 \%$, and $6.25\%$, respectively.
While a 25\% concentration exceeds the reported experimental averages of 3 to 15\%~\cite{yan2022perspective,Garnica2022}, it represents an upper bound for locally high defect densities. 
STM imaging has demonstrated that defect concentrations can fluctuate significantly across the sample~\cite{Garnica2022}. By exploring this concentrated limit, we can systematically investigate the evolution of the magnetic coupling and probe the defect-defect interactions that may arise under such local conditions.

The Mn-rich and intermixing models were selected to analyze the energy difference between FM and FiM configurations due to the presence of an additional Mn atom. The specific magnetic monolayer models used for these analyses are summarized in \cref{tab:configurations_ML} of the Appendix section. The energy difference between the FM and FiM configurations, normalized per SL per unit cell, is defined as
\begin{equation}
  \label{eq:deltaEML}
  \begin{aligned}
  \Delta E_{\mathrm{ML}} = E_{\mathrm{FM}} - E_{\mathrm{FiM}},
  \end{aligned}
\end{equation}
thus a positive (negative) value indicates the preference for the FiM (FM) ordering, respectively.

\begin{figure}[t!]
    \centering
    \includegraphics[width=0.45\textwidth]{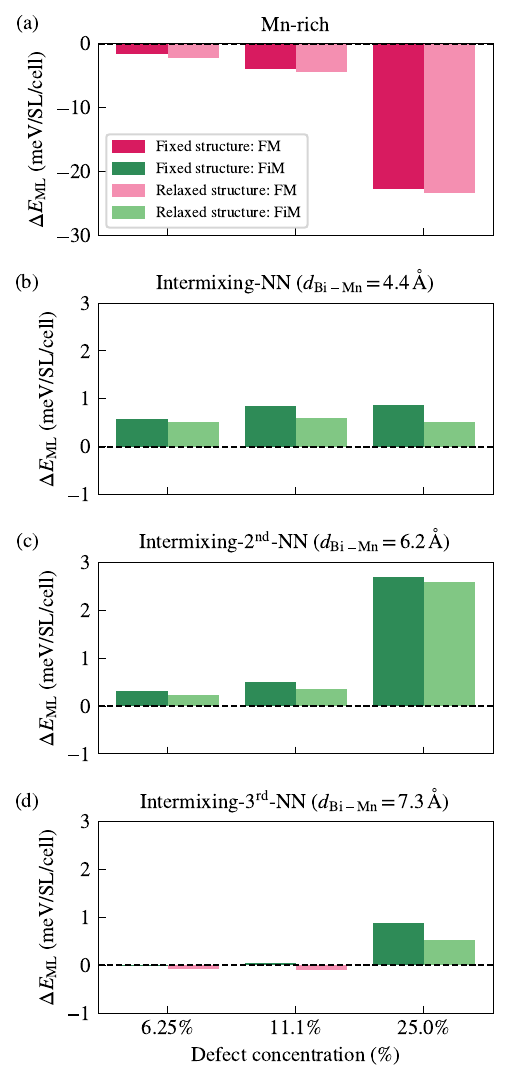} 
    \caption{Intralayer coupling as a function of the defect concentration of monolayers. (a) Mn-rich defect. Intermixing defect at a distance between Bi and Mn defects of (b) 4.4 \AA, (c) 6.2 \AA, and (d) 7.3 \AA.
    The dark bars show the energy difference between FM and FiM states for structures with fixed atomic positions. The light bars show the energy differences obtained after relaxing the
    internal atomic coordinates with fixed lattice parameters 
    for the FM and FiM configurations, respectively.
    }
    \label{fig:combinedML}
    \begin{minipage}[t]{0.45\textwidth}
      \phantomsubcaption
      \label{fig:Mn_rich_deltaEML}
  \end{minipage}
  \begin{minipage}[t]{0.45\textwidth}
      \phantomsubcaption
      \label{fig:IntNNdeltaEML}
  \end{minipage}
  \begin{minipage}[t]{0.45\textwidth}
      \phantomsubcaption
      \label{fig:IntNNNdeltaEML}
  \end{minipage}
  \begin{minipage}[t]{0.45\textwidth}
      \phantomsubcaption
      \label{fig:Int3dNNdeltaEML}
  \end{minipage}
\end{figure}

\Cref{fig:combinedML} shows the ground-state magnetic configurations for Mn-rich and intermixing defects in the monolayer. 
The dark bars represent the energy difference between the FM and FiM states,
calculated using the fixed ground-state atomic structure to decouple the magnetic exchange contribution from the structural effects.
The light bars incorporate internal relaxation, in which the atomic positions were optimized independently for each magnetic configuration while maintaining fixed lattice parameters. 

\Cref{fig:Mn_rich_deltaEML} illustrates
the result of the Mn-rich system.
There is a unique position for the Mn atom
due to the symmetry constraints. 
It is observed that, independently of the defect concentration and the structure configuration,
the magnetic moment of the Mn defects consistently aligns parallel to the magnetic moments of the Mn layer. 
Moreover, 
the total energy difference becomes further negative with increasing concentration of defects.
This trend is consistent with
the behavior observed in bulk Mn-rich systems, 
confirming that ferromagnetic coupling is enhanced 
by increasing \MnBi defect concentration.

\begin{figure}[t!]
      \centering
    \includegraphics[width=3.4in]{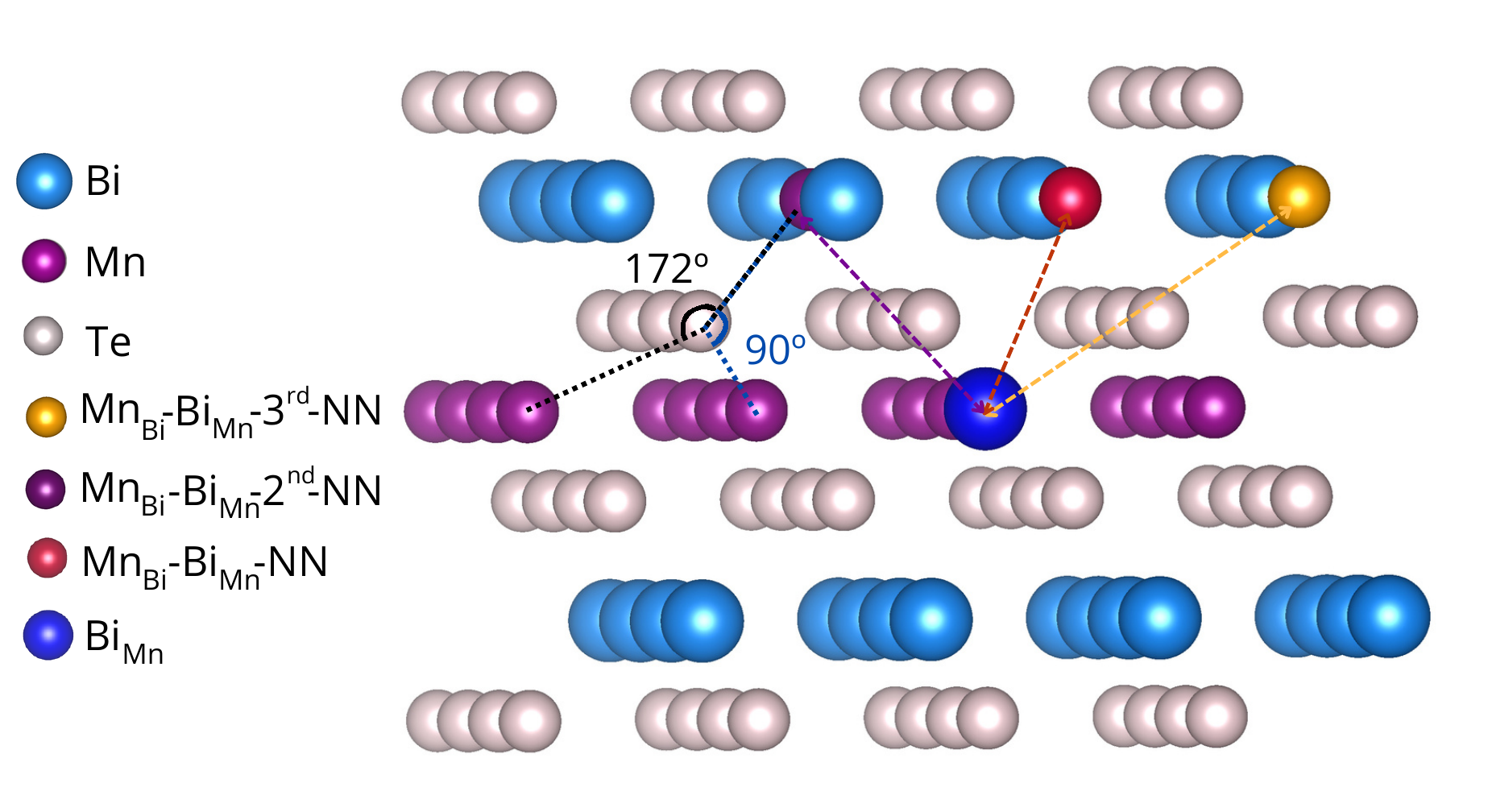}
    \captionsetup{justification=justified}
    \caption{\MBT monolayer including intermixing defects
    with different defective-site distances: nearest-neighbor (\MnBi-\BiMn-NN), second-nearest-neighbor (\MnBi-\BiMn-2\textsuperscript{nd}-NN), and third nearest-neighbor antisite defects (\MnBi-\BiMn-3\textsuperscript{rd}-NN). The bond angles of \MnBi-Te-Mn are approximately $90^\circ$ and $172^\circ$, measured relative to the nearest-neighbor and second-nearest-neighbor Mn atoms within the main layer.}
    \label{fig:intermix_sites}
\end{figure}

In the intermixing model, there are several possible site locations for the \MnBi and \BiMn.
The position of the Bi atom in the Mn layer is fixed, and the Mn is placed in three inequivalent Bi sites, from the nearest neighbor to the third nearest neighbor, as illustrated in \cref{fig:intermix_sites}.
Due to symmetry constraints, only the nearest and second-nearest-neighbors are achievable in the
$2\times2\times1$ supercell with the 25\% defect concentration. 
Consequently, the third-nearest-neighbor antisite defect at this concentration yields approximately the same result as that of the nearest neighbor, as demonstrated by \cref{fig:IntNNdeltaEML,fig:Int3dNNdeltaEML}.

To analyze the magnetic properties of the intermixing defect, the distance between the Bi and Mn defects, the concentration, and the local geometry of the defects are investigated,
as shown in
\cref{fig:IntNNdeltaEML,fig:IntNNNdeltaEML,fig:Int3dNNdeltaEML}.
Initially, we explored the effect of concentration on determining the preferred magnetic configuration while keeping the distance between the defects constant. 
Basically, the
Mn atom of the intermixing defect
exhibits a magnetic moment opposite to that of the Mn main layer, resulting in a FiM order that remains consistent regardless of concentration and distance.
This behavior is consistent with that of bulk intermixing systems.
Moreover, at concentrations of 6.25\% and 11.1\%, the ferrimagnetic strength decreases as the distance between the \MnBi and \BiMn defects increases.
Besides, when
the internal atomic coordinates are relaxed separately for the FM and FiM configurations with fixed lattice parameters, the results remain qualitatively consistent across nearly all cases.
The sole exception is the third-nearest neighbor intermixing, where the FM configuration appears more stable.
However, since $\Delta E_{\mathrm{ML}}$ for this model falls within the expected numerical precision, the preferred configuration remains uncertain. 
Furthermore, as illustrated in \cref{fig:Mn_rich_deltaEML,fig:IntNNNdeltaEML}, $\Delta E_{\mathrm{ML}}$ at 25\% is significantly enhanced, which could be attributed to finite-size interactions. Nonetheless, the predicted magnetic ground states remain robust across the investigated concentration range.

To move beyond total energy differences and identify the physical origin of these magnetic states, we analyze the novel exchange pathways induced by the \MnBi defect.
In accordance with the Goodenough-Kanamori-Anderson (GKA) rules, the $\text{Mn-Te-Mn}$ bond angle determines the sign of the Mn-Mn coupling through these pathways.
For the Mn 3\textit{d} half-filled shell, these rules state that a bonding angle of $90^\circ$ generally favors FM coupling, whereas a $180^\circ$ angle leads to an AFM interaction~\cite{Khomskii2014Oct,kanamori1959superexchange}.
As shown in \cref{fig:intermix_sites}, the defect creates a nearest-neighbor pathway exhibiting an angle of approximately $90^\circ$, which would promote FM coupling.
A second-nearest-neighbor pathway exhibiting an angle of approximately $172^\circ$ would favor antiparallel alignment of the magnetic moments. 
This geometrical distortion should then result in a competition between the FM and AFM superexchange components~\cite{Lai2021,advs.202402753}.

\subsection{Analysis of magnetic coupling in \MBT monolayer}

To further elucidate the stability of the FM ground state in the Mn-rich monolayer system and the FiM state in the intermixing-NN monolayer system, we mapped the magnetic interactions onto a Heisenberg Hamiltonian~\cite{HE2021107938}:  
\begin{equation}
    H = -\sum_{i \ne j} \left[ J^{\mathrm{iso}}_{ij} (\mathbf{S}_i \cdot \mathbf{S}_j) + \mathbf{S}_i \mathbf{J}^{\mathrm{ani}}_{ij} \mathbf{S}_j +\mathbf{D}_{ij} \cdot (\mathbf{S}_i \times \mathbf{S}_j) \right]
\end{equation}
where $\mathbf{S}_i$ and $\mathbf{S}_j$ are unit vectors representing the magnetic moments at sites $i$ and $j$. Here, $J^{\mathrm{iso}}_{ij}$ represents the isotropic exchange coupling, $\mathbf{J}^{\mathrm{ani}}_{ij}$ the anisotropic exchange tensor, and $\mathbf{D}_{ij}$ the Dzyaloshinskii-Moriya interaction (DMI) vector. 
In this convention, a positive (negative) $J^{\mathrm{iso}}$ denotes FM (AFM) coupling.
By extracting the exchange parameters ($J_{ij}$), we can quantitatively assess the competition between the local exchange pathways identified above.

We focus on a defect concentration of 11.1\% as a representative case. 

\cref{tab:energy_comparison} shows the qualitative agreement between VASP, SIESTA, and the mapped Heisenberg model regarding total energy differences, illustrating that our tight-binding model captures the principal physical picture, 
although a direct quantitative comparison of $\Delta E$ is precluded by the distinct theoretical frameworks underlying the three methods.
While the DMI and anisotropy are integral components to the Hamiltonian, they contribute negligibly to the total energy difference owing to cancellation effects, as demonstrated in \sfig{5} in the SM~\cite{supplement2025}.
Consequently, we identify the isotropic exchange $J^{\mathrm{iso}}$ as the primary driver of magnetic stability.
Detailed parameters and further analyses of all components are available in \sfigsRange{6}{11} in the SM~\cite{supplement2025}.

\begin{table}[t!]
\centering
\caption{Energy difference per septuple layer per unit cell, as defined by \cref{eq:deltaEML}. Negative and positive values denote FM and FiM ground states, respectively, in meV.}
\label{tab:energy_comparison}
\begin{ruledtabular}
\begin{tabular}{lccc}

System         & $\Delta E_{\text{VASP}}$ & $\Delta E_{\text{SIESTA}}$ & $\Delta E_{\text{TB2J}}$ \\ \hline 
Mn-rich        & $-3.91$                  & $-2.30$                    & $-1.58$                  \\
Intermixing-NN & $0.86$                   & $1.41$                     & $3.12$                   \\ 
\end{tabular}
\end{ruledtabular}
\end{table}

\begin{figure}[t]
    \centering
    \includegraphics[width=\linewidth]{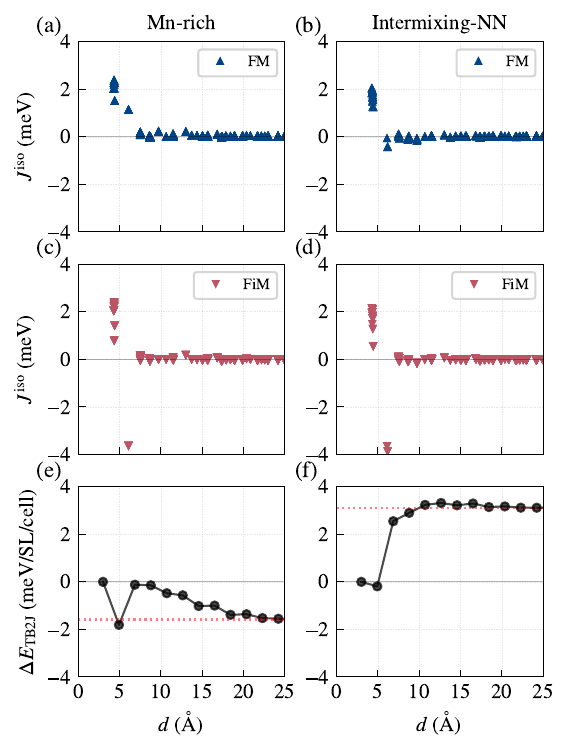}
  \caption{Isotropic exchange coupling ($J^{\mathrm{iso}}$) as a function of interatomic distance for the 11.1\% Mn-rich and intermixing-NN monolayer systems. Calculated parameters are shown for the (a), (b) FM and (c), (d) FiM configurations, while (e) and (f) display the cumulative sum of the $\Delta E_{\mathrm{TB2J}}$ components. Energy plateaus are reached between 20 and 25~{\AA} in (e) and around 10~{\AA} in (f), with interactions included up to 25~\AA. The horizontal red dotted line denotes the converged total energy value.}
\begin{minipage}[t]{0.49\textwidth}\phantomsubcaption\label{fig:MnrichFM}\end{minipage}
  \begin{minipage}[t]{0.49\textwidth}\phantomsubcaption\label{fig:Mn-richFiM}\end{minipage}
  \begin{minipage}[t]{0.49\textwidth}\phantomsubcaption\label{fig:IntNNFM}\end{minipage}
  \begin{minipage}[t]{0.49\textwidth}\phantomsubcaption\label{fig:IntNNFiM}\end{minipage}
  \begin{minipage}[t]{0.49\textwidth}\phantomsubcaption\label{fig:deltaETb2jMn-rich}\end{minipage}
  \begin{minipage}[t]{0.49\textwidth}\phantomsubcaption\label{fig:deltaETb2jintermix}\end{minipage}
  \label{fig:jiso}
\end{figure}

\Cref{fig:jiso} illustrates the isotropic exchange coupling as a function of the interatomic distance for the Mn-rich and intermixing-NN systems across both FM and FiM configurations. 
Both systems exhibit strong FM nearest-neighbor interactions at $d \approx 4.3$~{\AA}, with $J^{\mathrm{iso}}$ values ranging from $0.81$ meV to $2.40$ meV. The observed dispersion in $J^{\mathrm{iso}}$ values arises from defect-induced symmetry breaking, which creates a distribution of inequivalent Mn–Mn bond lengths. 

The sign of the second-nearest-neighbor interactions at $d \approx 6.1$~{\AA} depends on the system.
In the Mn-rich system, $J^{\mathrm{iso}}$ is 1.13 meV for the FM configuration, but shifts to $-3.62$~meV in the FiM configuration, as depicted in \cref{fig:MnrichFM} and \cref{fig:Mn-richFiM}. 
Conversely, in the intermixing system, the antiparallel coupling is robust in both FM and FiM configurations, reaching values of $-0.44$~meV and $-3.85$~meV, as illustrated in \cref{fig:IntNNFM,fig:IntNNFiM}. 

The electronic nature of the defect-induced phase further dictates the spatial range and magnitude of these interactions, as reflected by the cumulative sum of the exchange components as a function of distance in \cref{fig:deltaETb2jMn-rich,fig:deltaETb2jintermix}.
The Mn-rich system exhibits long-range exchange interactions that contribute to the total magnetic energy up to $25$~{\AA}, as illustrated in \cref{fig:deltaETb2jMn-rich}.
In this case, the cumulative density of short-range and long-range FM pairs outweighs the second-nearest-neighbor antiparallel components, stabilizing the FM ground state. 
This extended magnetic coupling is mediated by the metallic character of the Mn-rich system, as verified by the projected density of states (PDOS) in \sfig{3} of the SM~\cite{supplement2025}.
In contrast, for the intermixing-NN system, the short-range interactions dominate the exchange energy landscape, with the cumulative $\Delta E_{\mathrm{TB2J}}$ converging rapidly within 10~{\AA}, as shown in \cref{fig:deltaETb2jintermix}.
Here, the $\text{Bi}_{\text{Mn}}$ defect diminishes the number of first-nearest-neighbor FM pairs, allowing the second-nearest-neighbor interactions to prevail.
This stabilizes the FiM alignment via a mechanism consistent with the GKA superexchange rules. This is supported by the semiconducting nature of the system, which is confirmed by the PDOS in \sfig{4} of the SM~\cite{supplement2025}.

\section{Conclusion}

Focusing on intrinsic defects, the magnetic ground states of defective bulk and monolayer \MBT systems have been investigated theoretically using state-of-the-art $ab~initio$ calculations.
In bulk systems, the Mn-vacancy defect enhances the ferromagnetic coupling,
the \BiMn and \MnBi-\BiMn intermixing defects preserve the interlayer antiferromagnetic coupling,
while \MnBi defects induce an antiferromagnetic-to-ferromagnetic transition of the interlayer coupling at increasing concentration.  
In the monolayer, Mn-rich configurations consistently favored ferromagnetic alignment, whereas intermixing defects favored ferrimagnetic states. 
Mapping the first-principles results onto a Heisenberg Hamiltonian reveals that the magnetic energy differences are dominated by isotropic exchange interactions.
Our results identify defects as a tuning knob for the magnetic ground state of \MBT, 
confirming that Mn vacancies and Mn excesses provide an effective route to drive an antiferromagnetic-to-ferromagnetic transition.
By correlating defect-induced electronic modifications to exchange mechanisms, this work establishes a theoretical foundation for interpreting diverse experimental observations and for the magnetization control of \MBT.

% --- THE APPENDIX STARTS HERE ---
\appendix
\renewcommand{\thesection}{}
\section{Atomic models}

% Redefine the table numbering format
\renewcommand{\thetable}{A\arabic{table}}
% Reset the counter if you want it to start from I again
\setcounter{table}{0}

% --- TABLE 1: BULK MODELS (Two-column span) ---
\begin{table}[th]
\centering
\small
\caption{Atomic structure configurations for bulk $\text{MnBi}_2\text{Te}_4$ defect models. Each model shows the various magnetic orderings considered in a $2\times2\times1$ supercell. Structural parameters and relaxation details are provided in the Methods section.}
\label{tab:configurations_bulk}

%\begin{tabular}{lcccc} 

\renewcommand{\arraystretch}{1.2} 

% Use tabular* with \columnwidth
\begin{tabular*}{\columnwidth}{@{\extracolsep{\fill}} lcccc @{}}
\hline \hline
Model & FM & AFM & AFiM & FiM \\ \midrule
% --- Pristine ---
Pristine & 
\raisebox{-.5\height}{\includegraphics[height=1.3in]{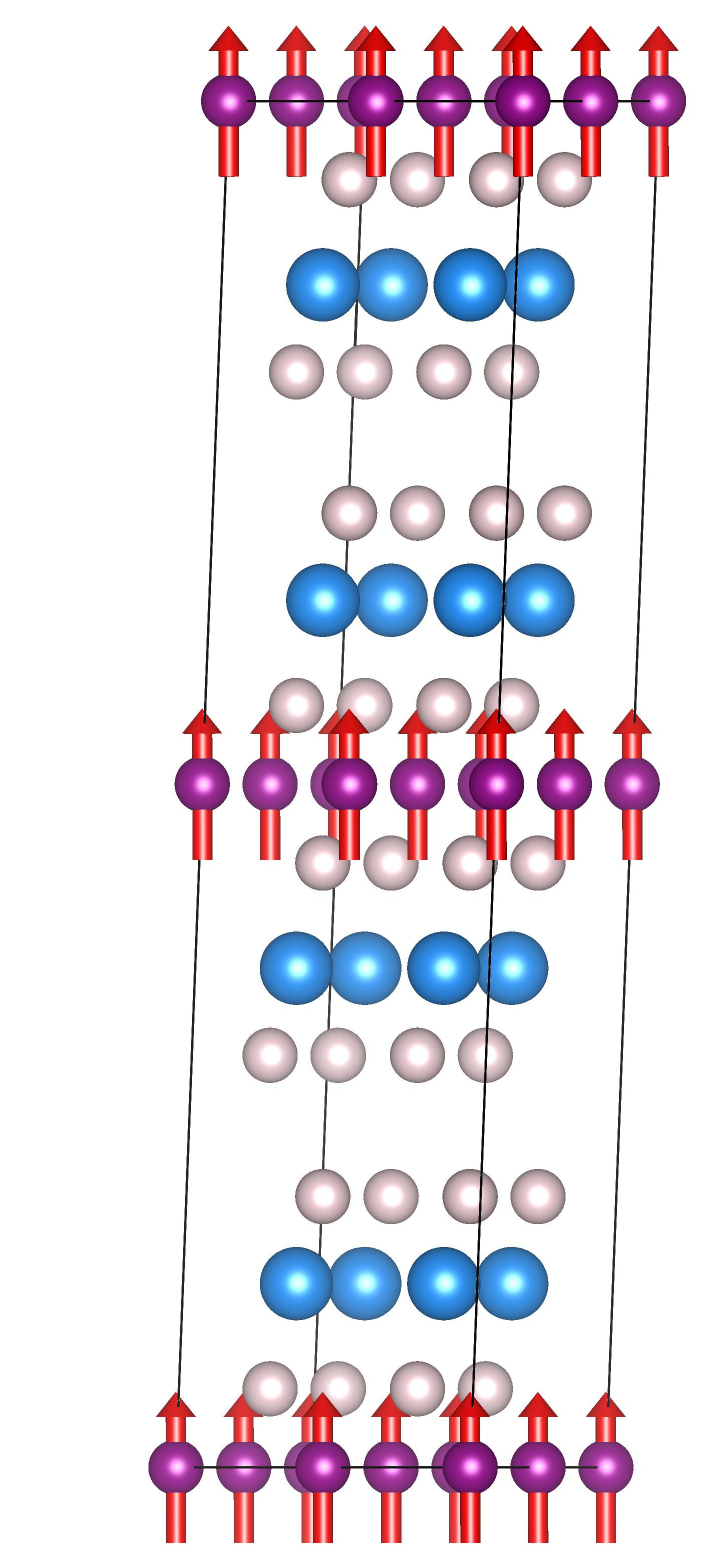}} & 
\raisebox{-.5\height}{\includegraphics[height=1.3in]{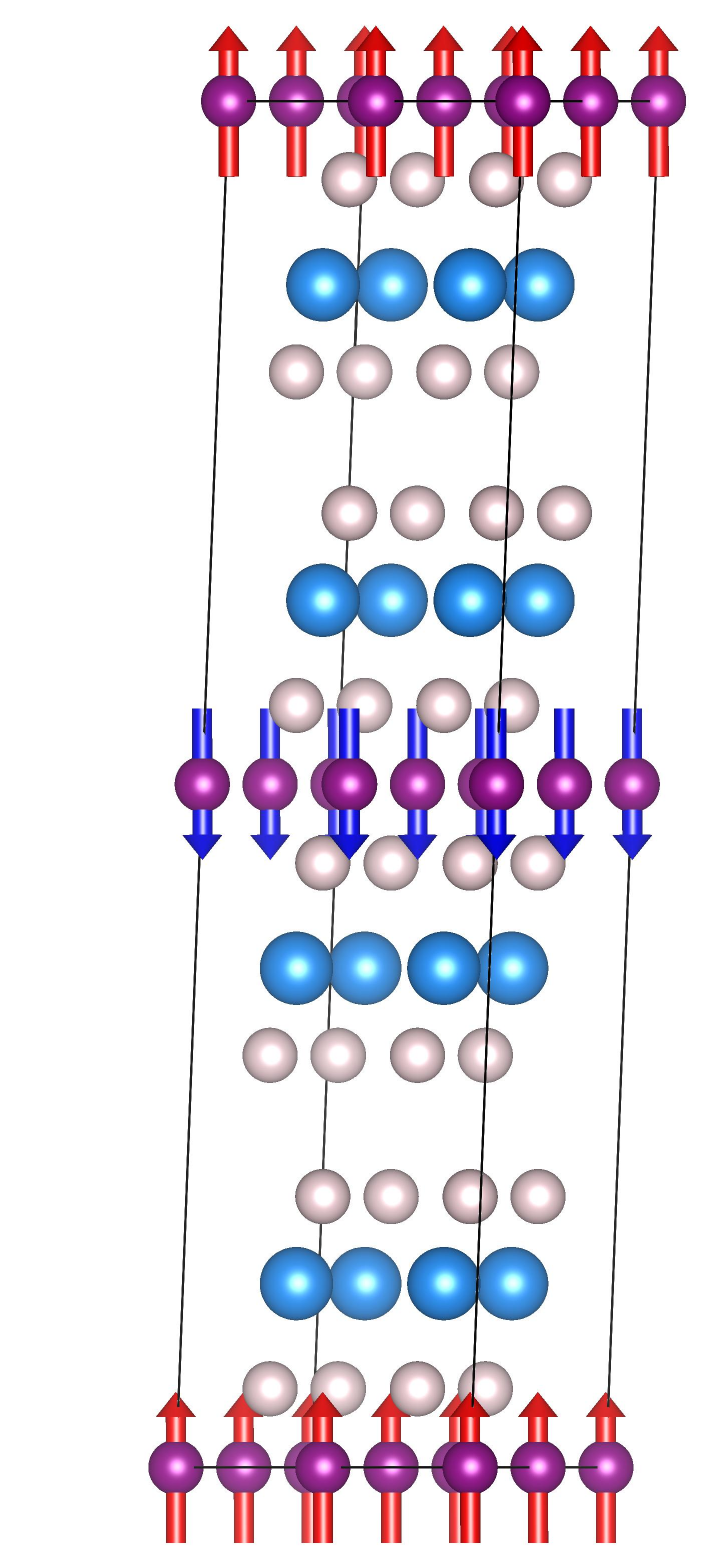}} & 
--- & --- \\ \midrule
% --- Mn-vacancy ---
Mn-vacancy & 
\raisebox{-.5\height}{\includegraphics[height=1.3in]{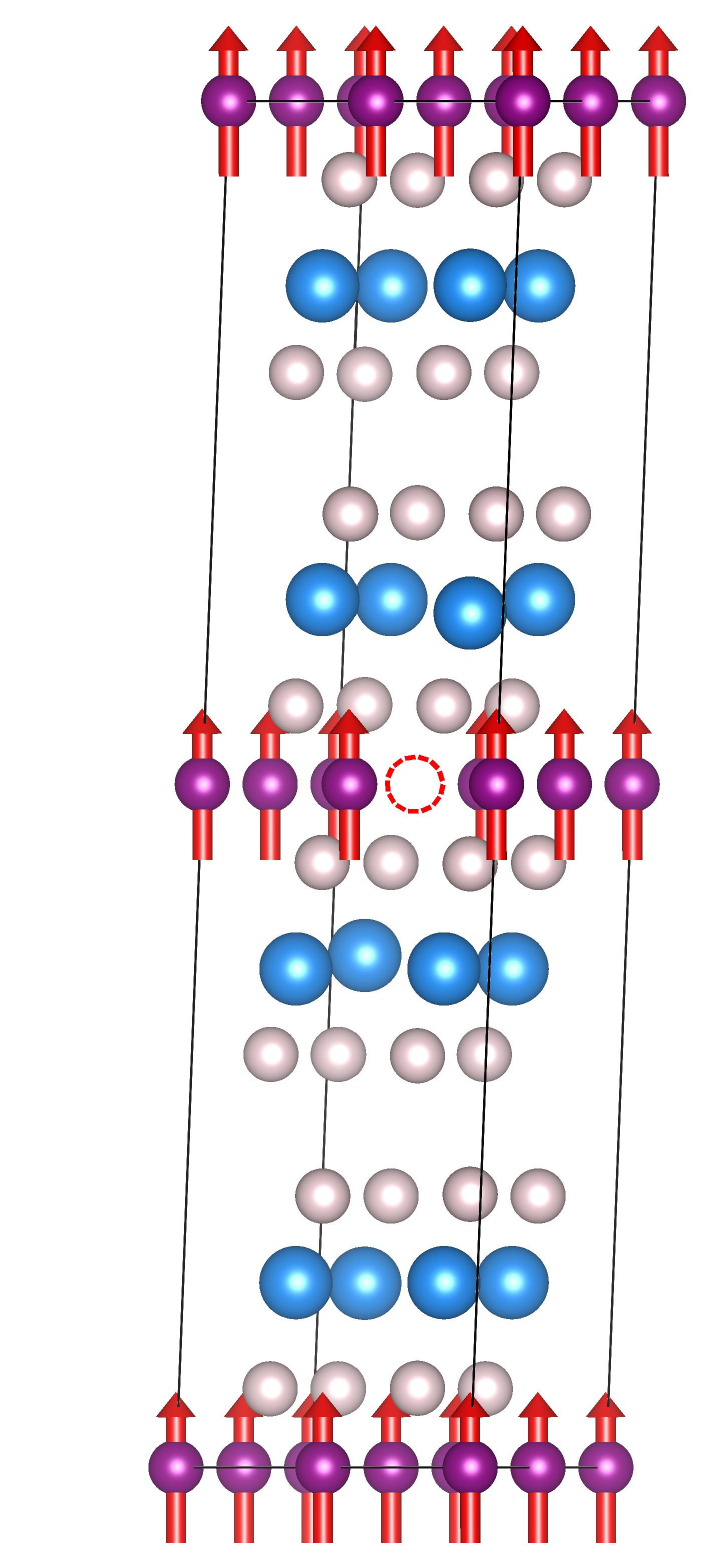}} & 
\raisebox{-.5\height}{\includegraphics[height=1.3in]{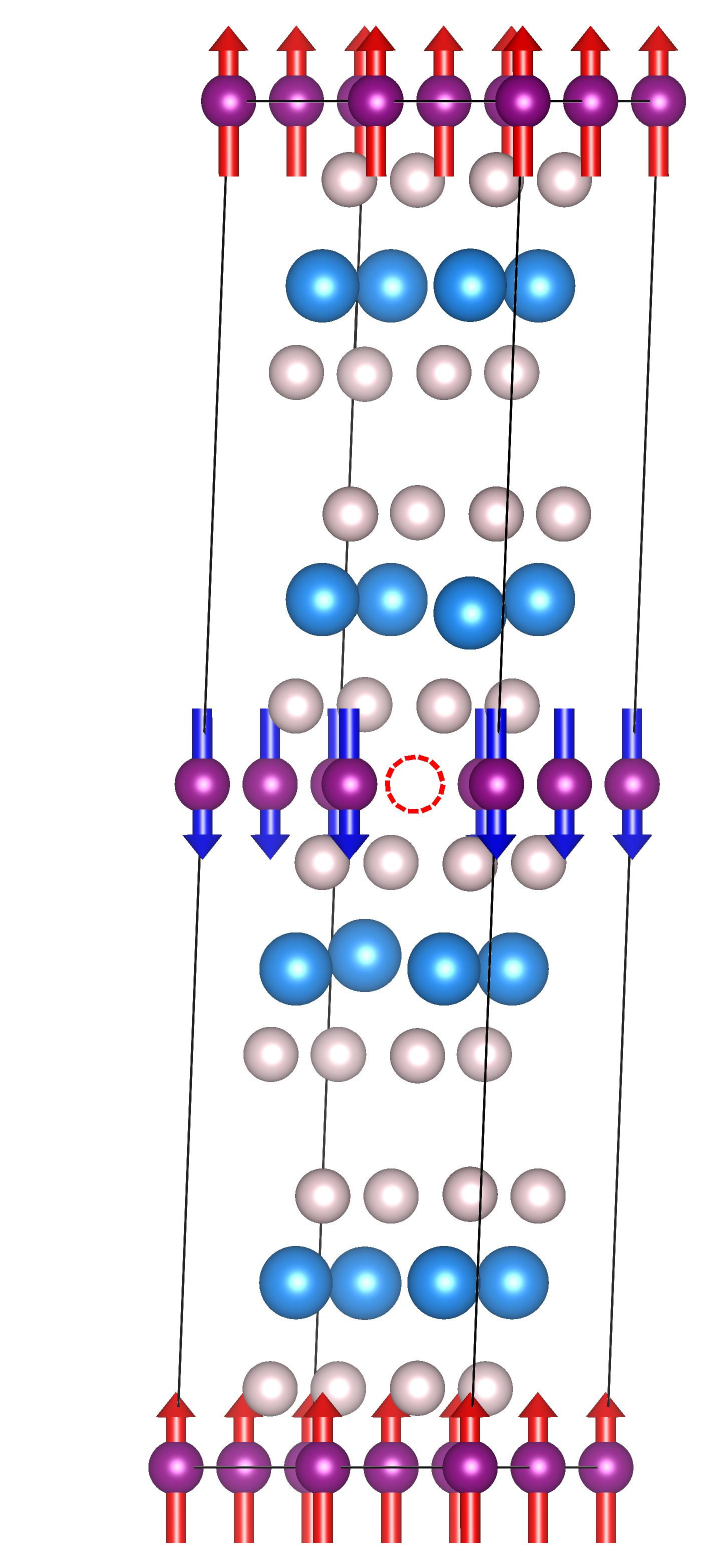}} & 
--- & --- \\ \midrule
% --- Bi-rich ---
Bi-rich & 
\raisebox{-.5\height}{\includegraphics[height=1.3in]{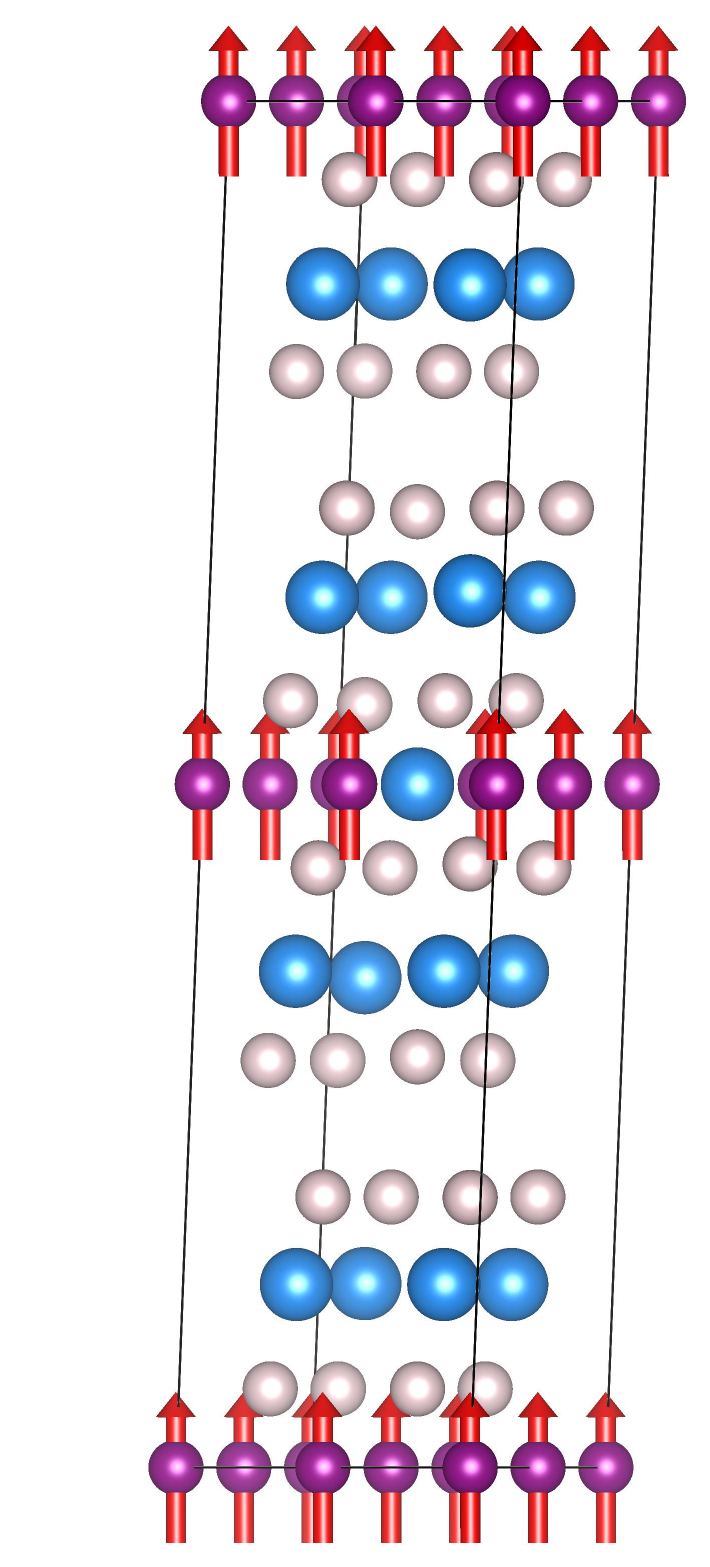}} & 
\raisebox{-.5\height}{\includegraphics[height=1.3in]{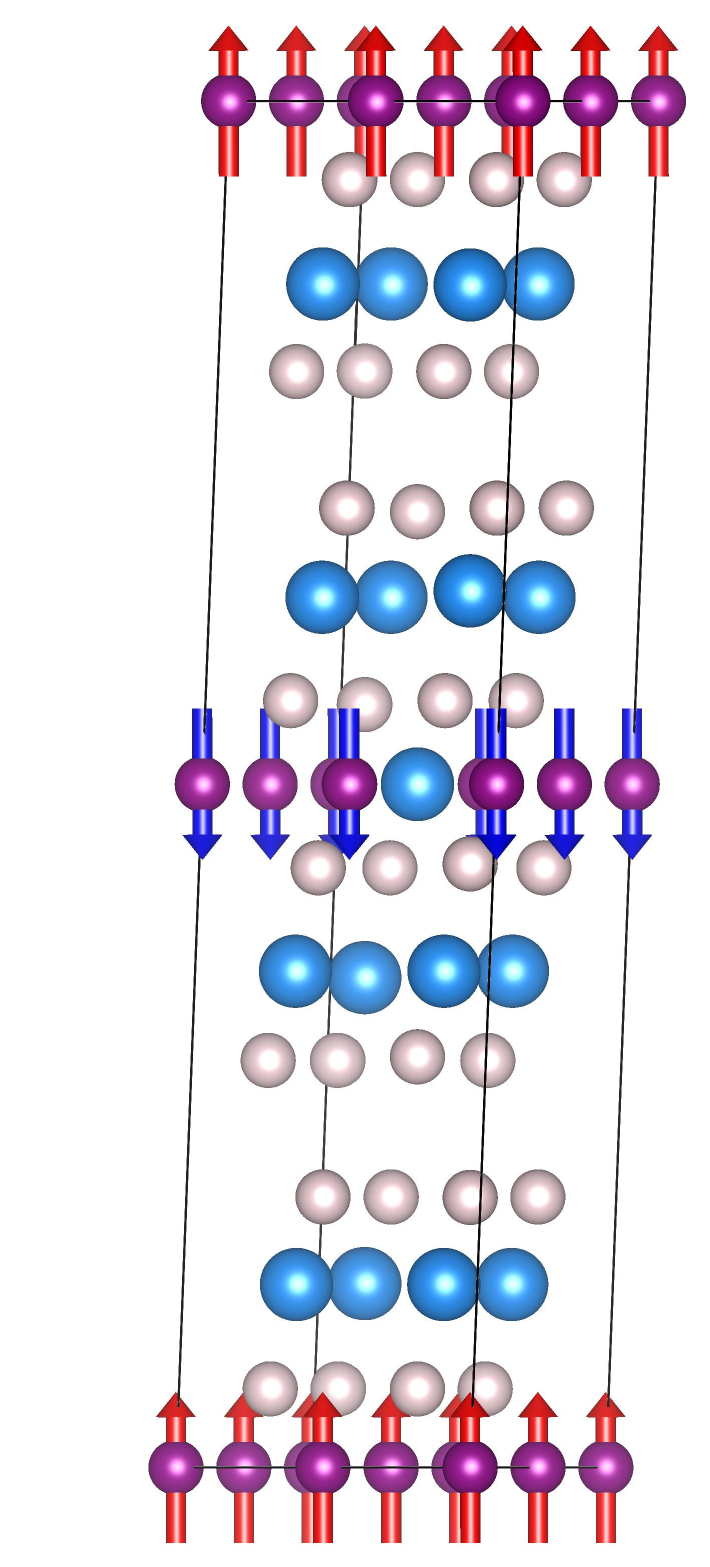}} & 
--- & --- \\ \midrule
% --- Mn-rich ---
Mn-rich & 
\raisebox{-.5\height}{\includegraphics[height=1.3in]{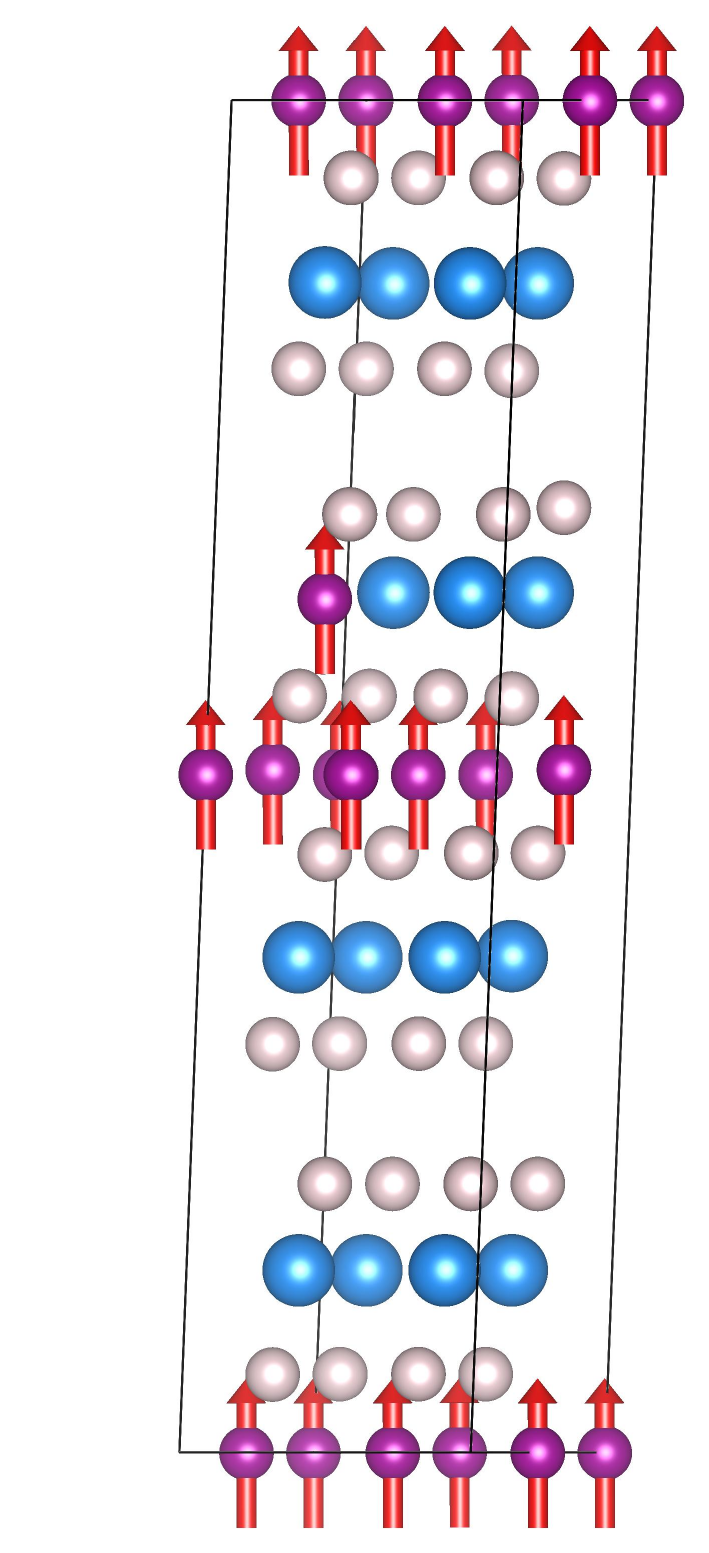}} & 
\raisebox{-.5\height}{\includegraphics[height=1.3in]{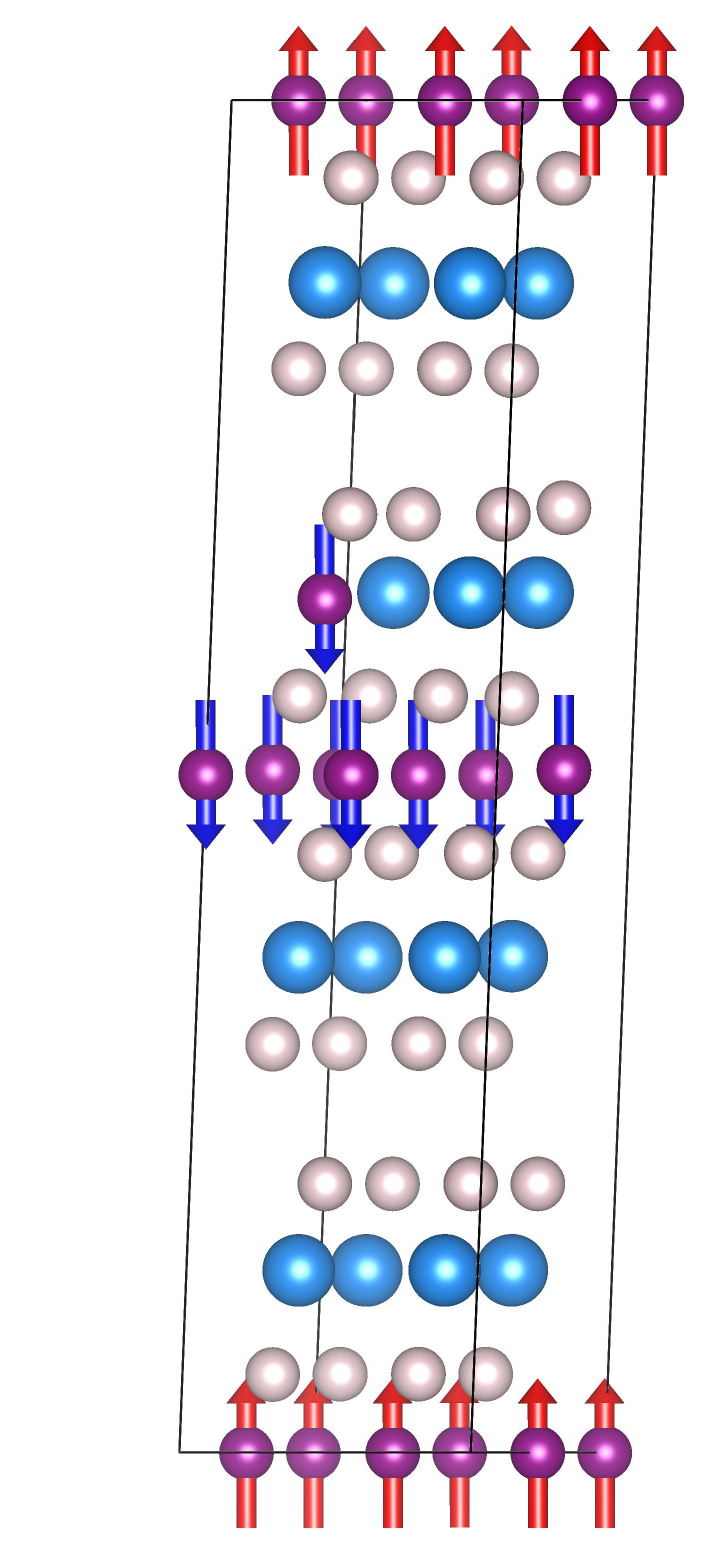}} & 
\raisebox{-.5\height}{\includegraphics[height=1.3in]{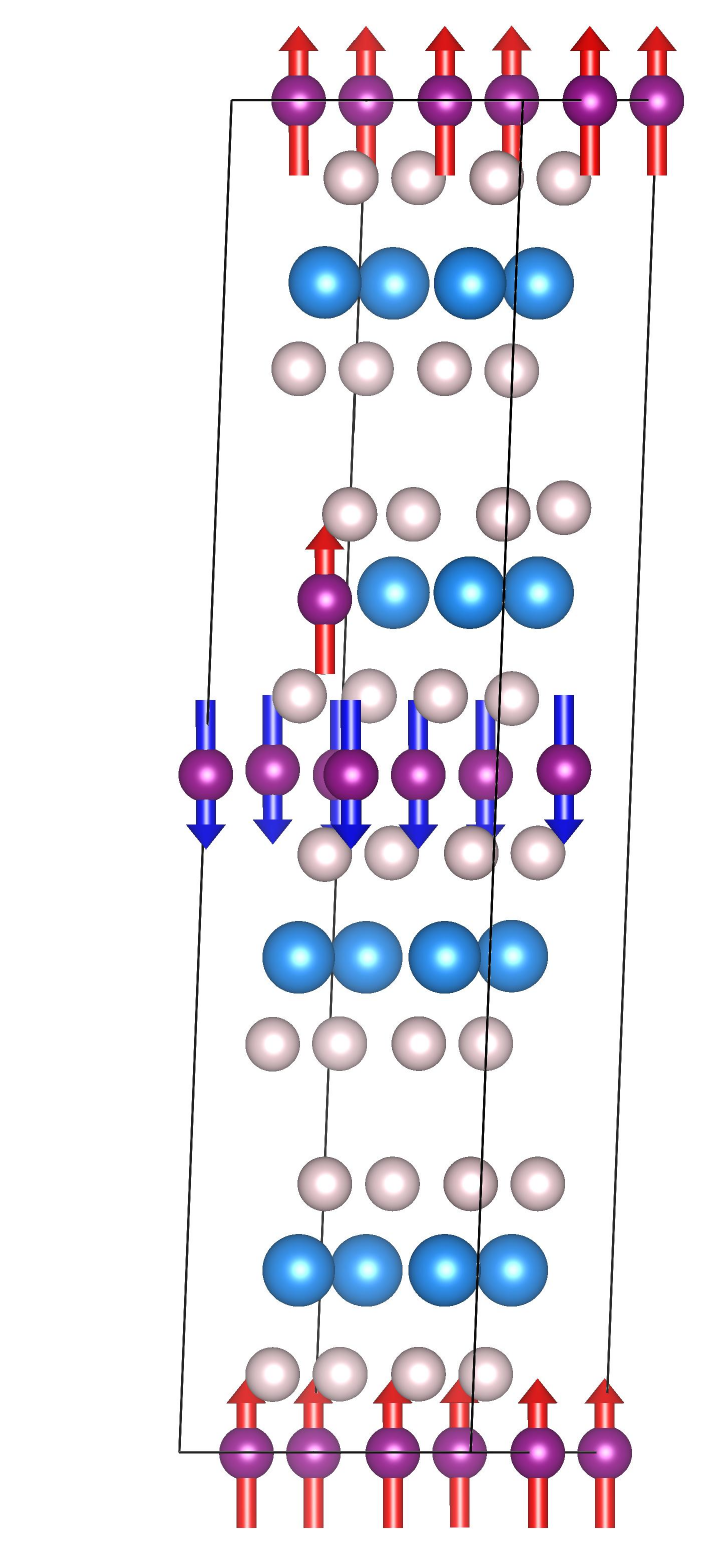}} & 
\raisebox{-.5\height}{\includegraphics[height=1.3in]{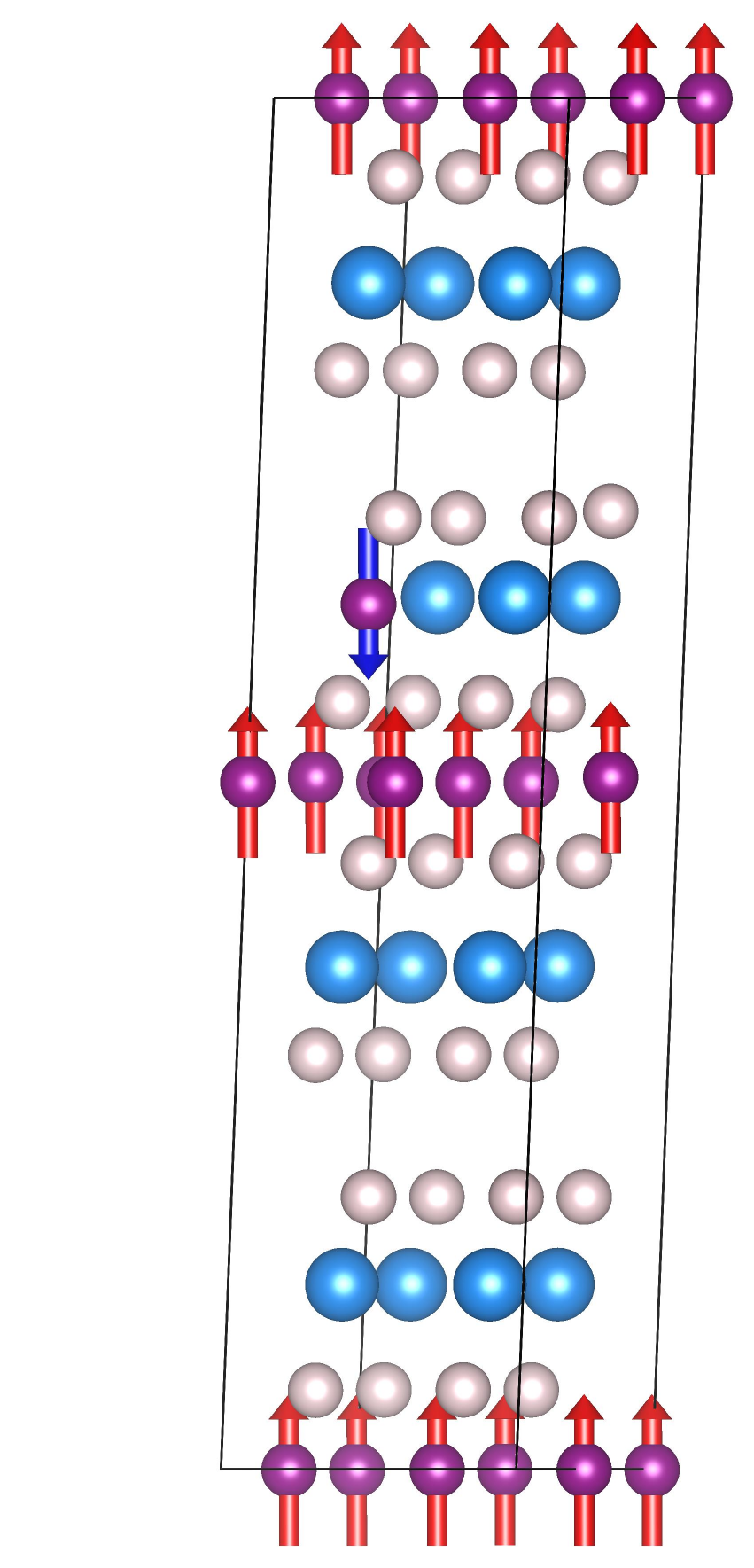}} \\ \midrule
% --- Intermixing ---
Intermixing & 
\raisebox{-.5\height}{\includegraphics[height=1.3in]{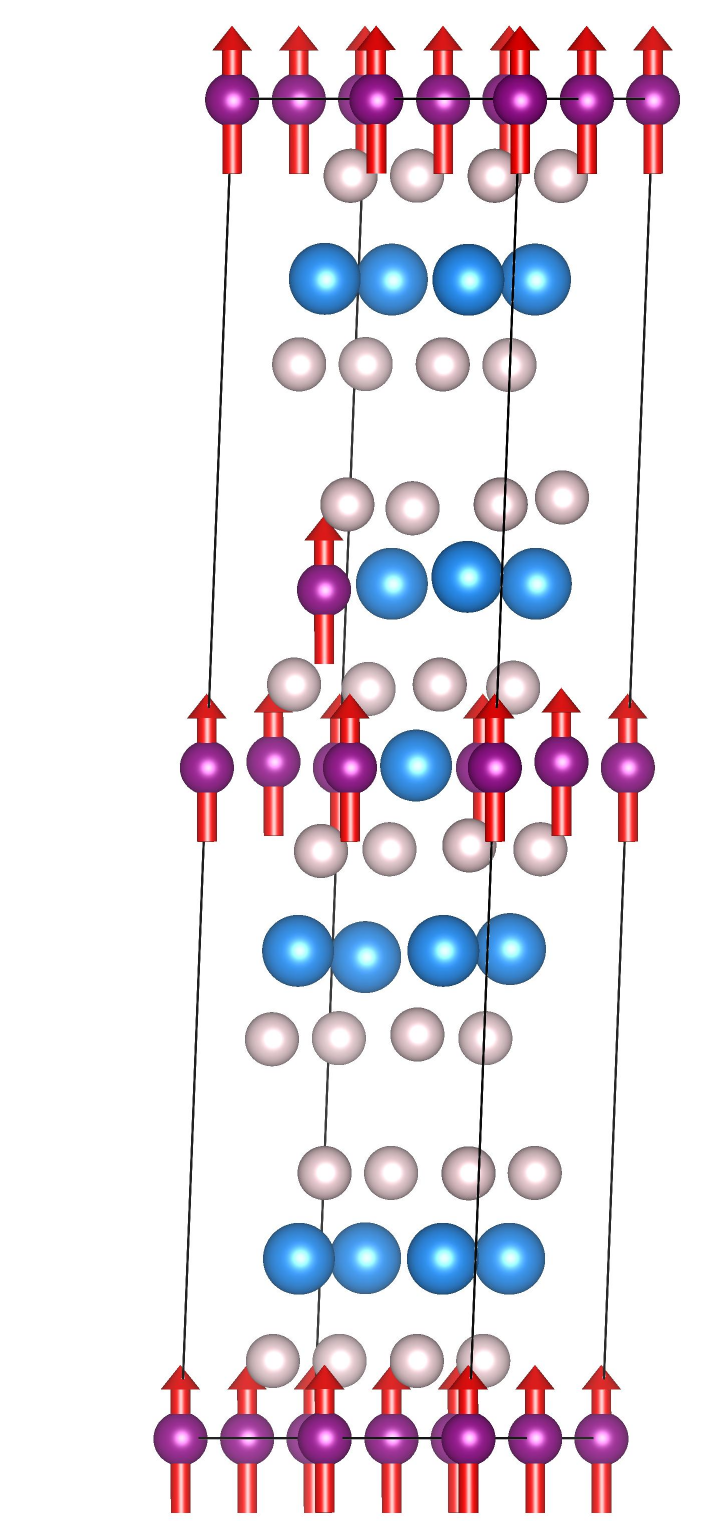}} & 
\raisebox{-.5\height}{\includegraphics[height=1.3in]{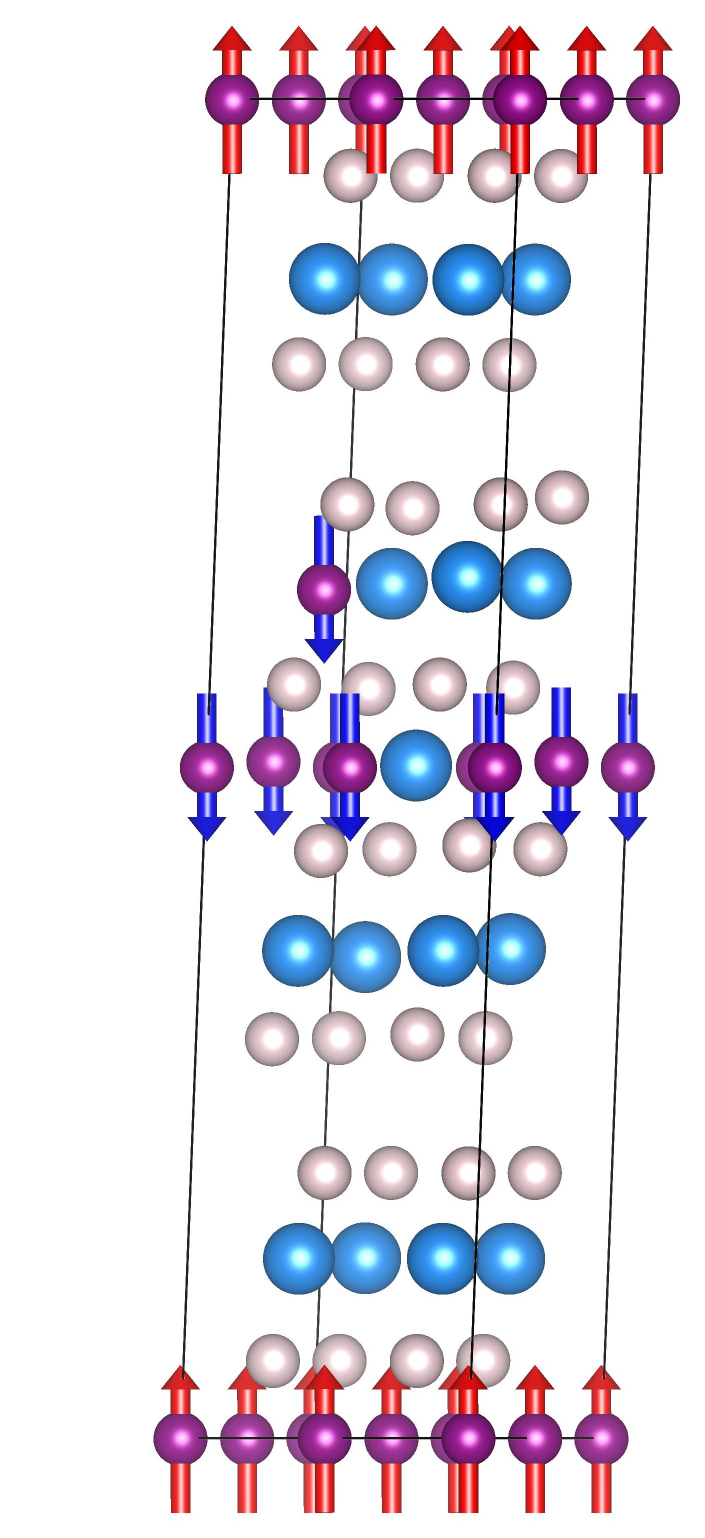}} & 
\raisebox{-.5\height}{\includegraphics[height=1.3in]{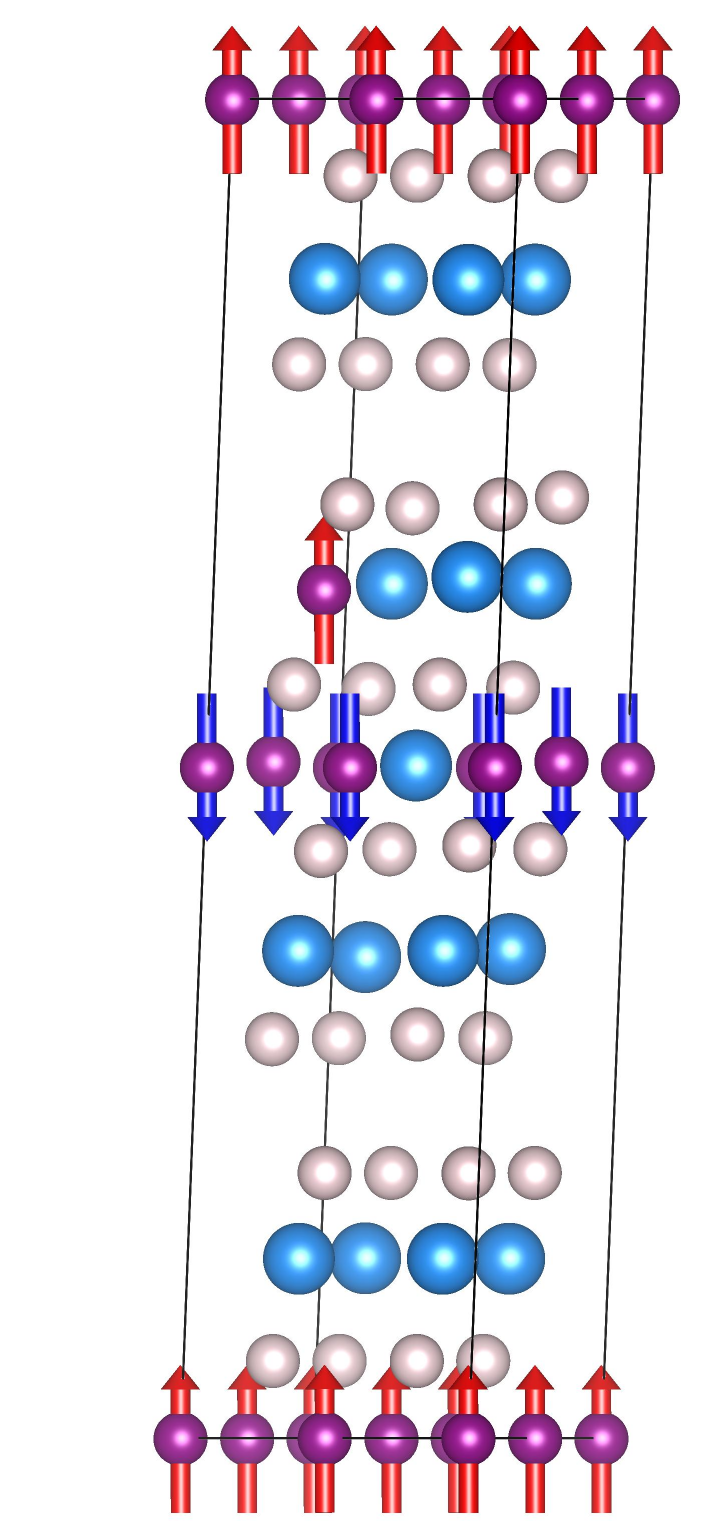}} & 
\raisebox{-.5\height}{\includegraphics[height=1.3in]{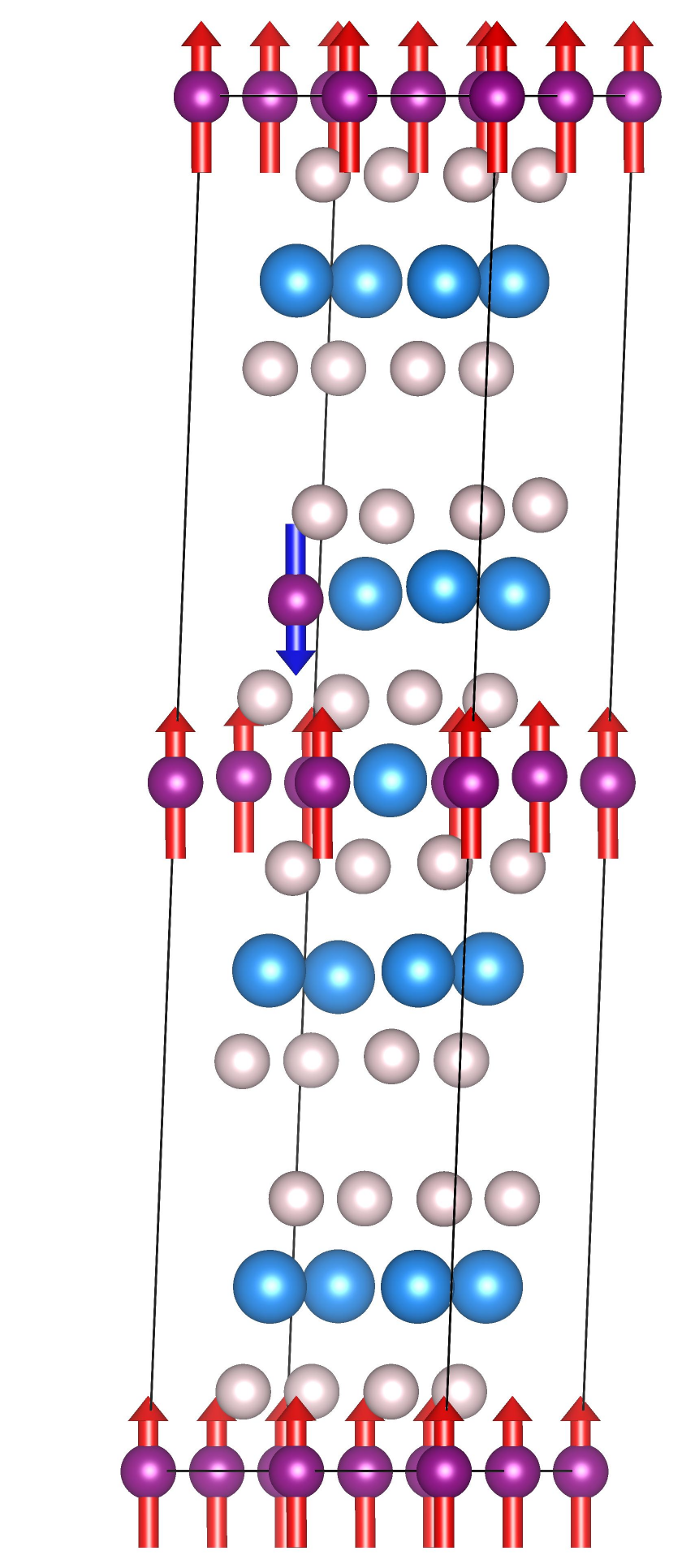}} \\ 
\hline \hline 
\end{tabular*}

\end{table}

\newpage
% --- TABLE 2: MONOLAYER MODELS (Two-column span) ---
\begin{table}[t]
\centering
\small
\caption{Atomic structure configurations for $\text{MnBi}_2\text{Te}_4$ monolayer defect models in a $2\times2\times1$ supercell. Structural parameters and relaxation details are provided in the Methods section.}
\label{tab:configurations_ML}

\renewcommand{\arraystretch}{1.2} 

% Use tabular* with \columnwidth
\begin{tabular*}{\columnwidth}{@{\extracolsep{\fill}} lcc @{}}
\hline \hline
Model & FM & FiM \\ \midrule
% --- ML Pristine ---
Pristine & 
\raisebox{-.5\height}{\includegraphics[height=1.3in]{pristine-FM-2x2.pdf}} & 
--- \\ \midrule
% --- ML Mn-rich ---
Mn-rich & 
\raisebox{-.5\height}{\includegraphics[height=1.3in]{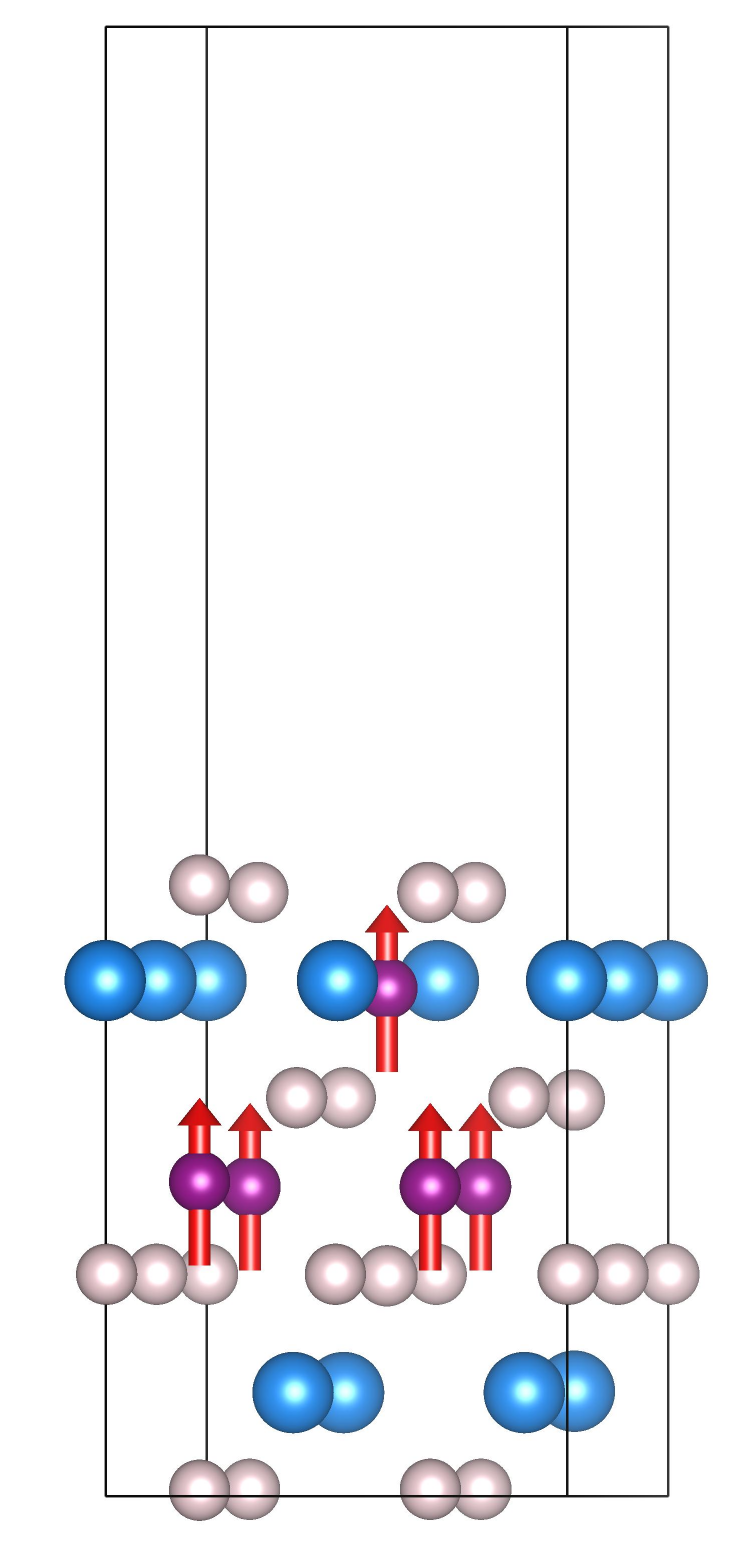}} & 
\raisebox{-.5\height}{\includegraphics[height=1.3in]{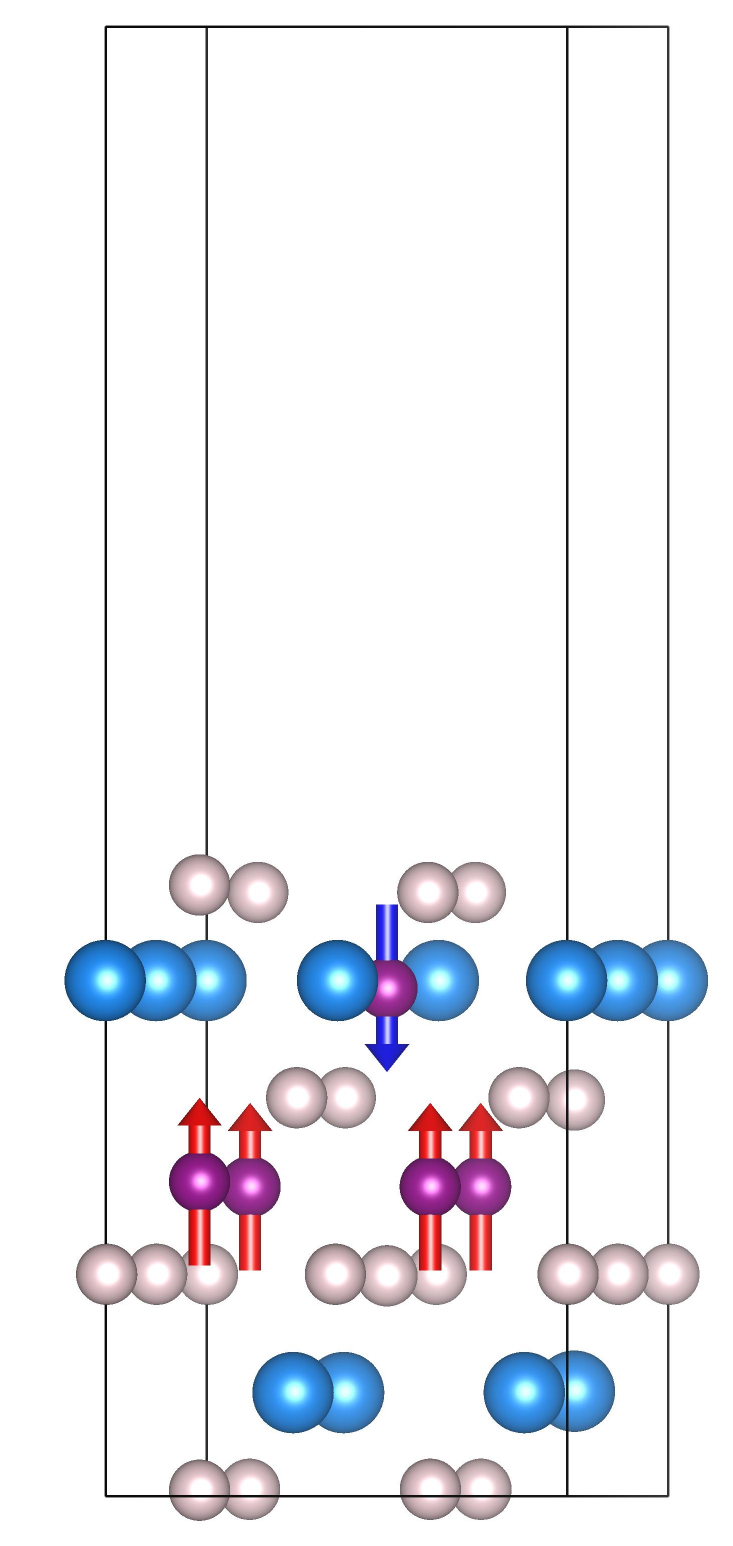}} \\ \midrule
% --- ML Intermixing ---
\begin{tabular}[c]{@{}l@{}} Intermixing-NN  \end{tabular} & 
\raisebox{-.5\height}{\includegraphics[height=1.3in]{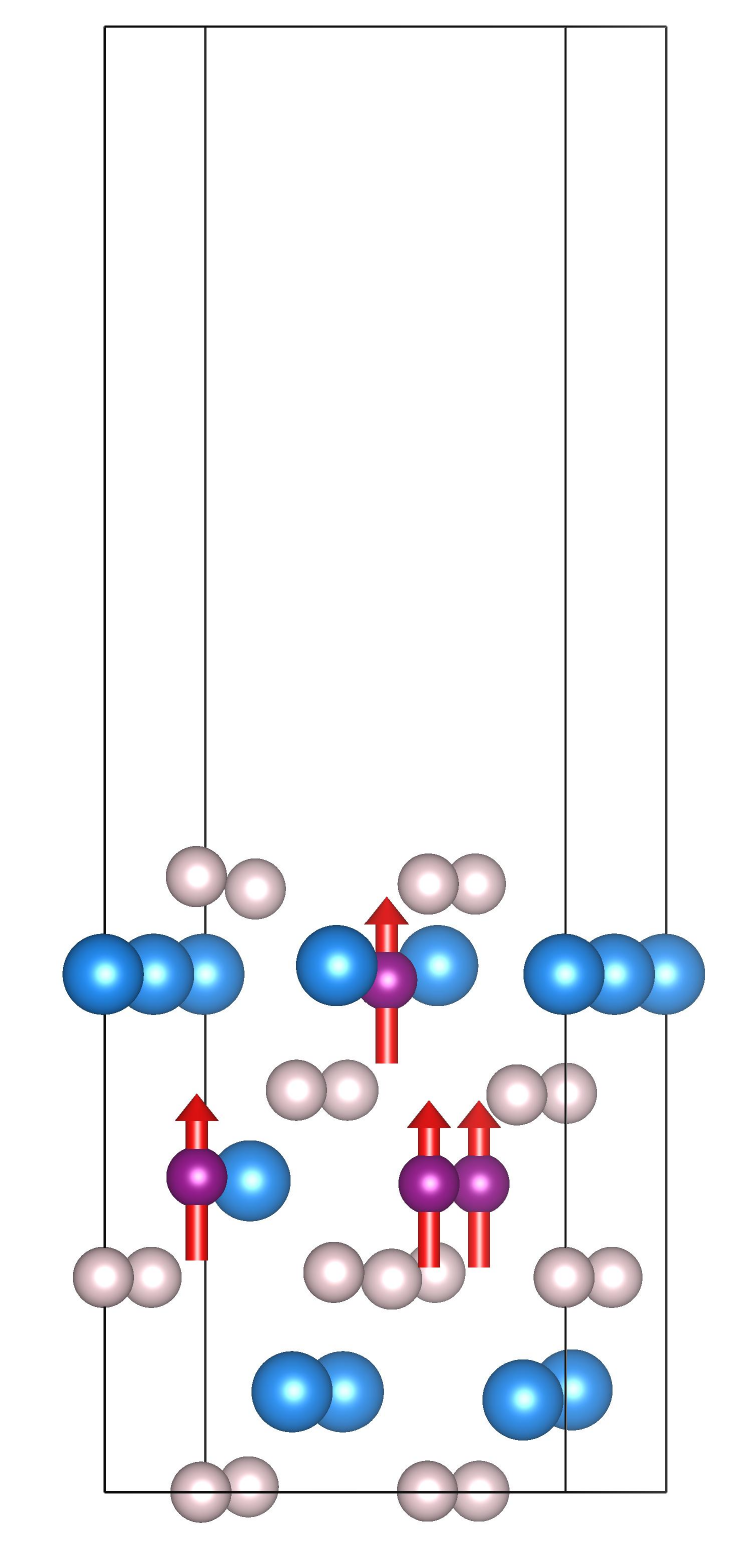}} & 
\raisebox{-.5\height}{\includegraphics[height=1.3in]{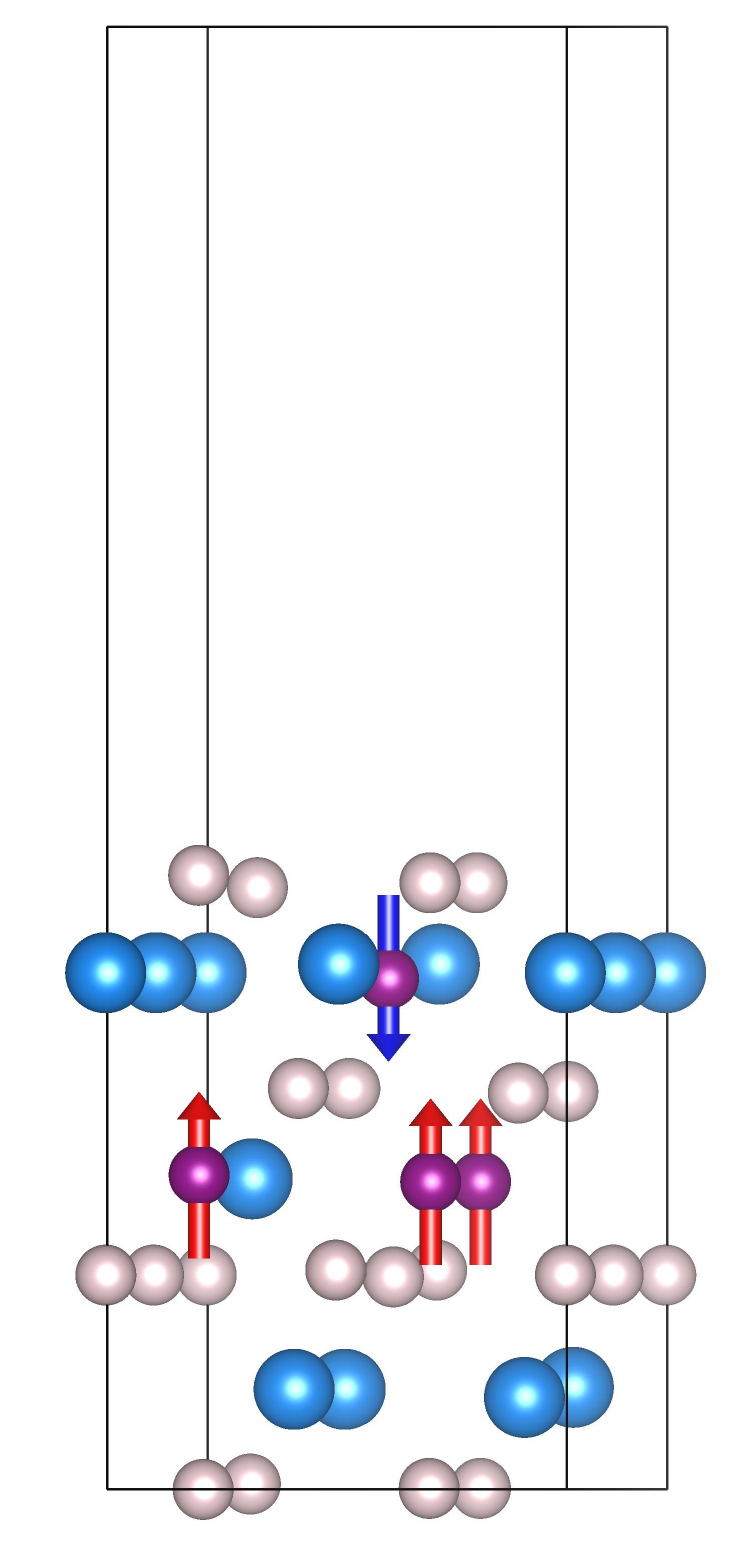}} \\ 
\hline
\hline
\end{tabular*}
\end{table}

\section*{ACKNOWLEDGMENTS}

The authors acknowledge financial support from the EOS project “CONNECT” (No. 40007563), from the Fédération Wallonie-Bruxelles through the ARC Grant “DREAMS” (N° 21/26-116), and from the F.R.S.-FNRS through the research project “MOIRÉ” (No. T.029.22F).
Computational resources were provided by the supercomputing facilities of the Université catholique de Louvain (CISM) and the Consortium des Équipements de Calcul Intensif en Fédération Wallonie Bruxelles (CÉCI) funded by the Fonds de la Recherche Scientifique de Belgique (F.R.S.-FNRS) under the convention No. 2.5020.11.

\section*{DATA AVAILABILITY}

The data that support the findings of this study are available from the corresponding author upon reasonable request.

\bibliography{ref.bib}
\end{document}

\title{Supplemental Material: \textit{Ab initio} study of magnetism in pristine and defective \MBT}%

\author{Ana Beatriz Pedro Fontes}
% \email{ana.pedrofontes@uclouvain.be}
\affiliation{Institute of Condensed Matter and Nanosciences (IMCN), Université catholique de 
Louvain (UCLouvain), B-1348, Louvain-la-Neuve, Belgium}

\author{Jiaqi Zhou}
% \email{jiaqi.zhou@uclouvain.be}
\affiliation{Institute of Condensed Matter and Nanosciences (IMCN), Université catholique de 
Louvain (UCLouvain), B-1348, Louvain-la-Neuve, Belgium}

\author{Simon M.-M. Dubois}
% \email{simon.dubois@uclouvain.be}
\affiliation{Institute of Condensed Matter and Nanosciences (IMCN), Université catholique de 
Louvain (UCLouvain), B-1348, Louvain-la-Neuve, Belgium}

\author{Jean-Christophe Charlier}
% \email{jean-christophe.charlier@uclouvain.be}
\affiliation{Institute of Condensed Matter and Nanosciences (IMCN), Université catholique de 
Louvain (UCLouvain), B-1348, Louvain-la-Neuve, Belgium}

\maketitle
\vspace{-15pt}  % Small space between title and TOC
\begingroup
\setlength{\parskip}{0pt}
\setlength{\itemsep}{0pt}
\tableofcontents
\vspace{-15pt} 
\endgroup

%\date{\today}%
\newpage

\newpage
\section{Cutoff energy convergence}
\begin{figure}[ht]
    \centering
    \includegraphics[width=\textwidth]{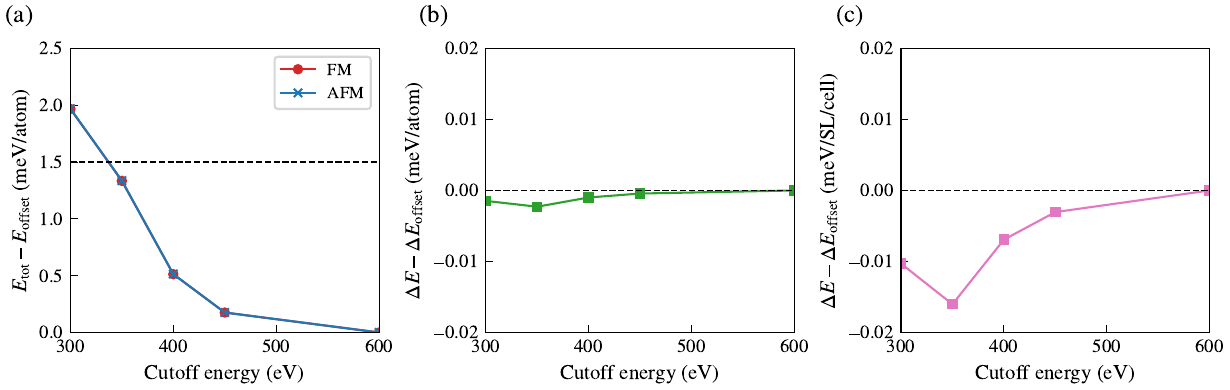}
    \caption{Convergence tests for the plane-wave cutoff energy. 
    Calculations performed for pristine bulk \MBT with SOC and $U=4$ eV. 
    (a) Total energy per atom for FM and AFM configurations 
    relative to the energy at 600 eV ($E_{\text{offset}}$). The dashed line indicates that the energy 
    convergence is within 1.5 meV/atom at 350 eV. 
    (b) Magnetic energy difference ($\Delta E = E_{\text{FM}} - E_{\text{AFM}}$) per atom and 
    (c) per septuple layer (SL) per unit cell, relative to the reference value at 600 eV 
    ($\Delta E_{\text{offset}}$). The stabilization in panels (b) and (c) demonstrates error cancellation between FM and AFM, ensuring that the relative magnetic stability is converged 
    to within 0.02 meV/SL/cell at the chosen 350 eV cutoff.}
    \label{fig:combined_ecutconvergence}
\end{figure}

\section{DFT+U}

Density functional theory (DFT), at the levels of local density approximation (LDA) or generalized gradient approximation (GGA), frequently encounters challenges in precisely characterizing the electronic properties of systems that contain partially filled \textit{d} or \textit{f} orbitals, including transition metals and rare earth elements. This inadequacy is primarily attributable to self-interaction errors (SIE), wherein DFT fails to completely cancel the interaction of an electron with itself, thereby resulting in an underestimation of localization effects~\cite{Dudarev1998Jan}. DFT+U is a widely used method due to its reasonable computational cost and the ability to improve the description of transition-metal systems. The Hubbard parameter U represents the Coulomb energy cost of placing two electrons on the same atomic site. The DFT+U incorporates an additional Coulomb potential term between the d and f orbitals at the same site, correcting the SIE inherent in standard DFT~\cite{Ricca2020Jun}.

An initial analysis was carried out to evaluate the stability of the magnetic ordering with different Hubbard U values by fully optimizing the crystal structure for each U value. Table \ref{tab:U_value_pristine} presents the band gap value calculated with PBE, the percentage error in the lattice parameters compared to the experimental values~\cite{Lee2013Jun}, and the total energy difference between the AFM and FM phases. The results show that the magnetic ordering remained qualitatively consistent for U values ranging from 0 to 5, with an increase in the Hubbard U leading to a decrease in the AFM interlayer coupling. A U value equal to 6 eV leads to an incorrect prediction of the magnetic ground state. In comparison, a U equal to 0 eV implies an indirect character of the band gap contrary to the experimental findings. 

The experimentally determined bulk band gap from photoemission measurements is estimated to be around 0.2~eV~\cite{Otrokov2019Dec}. While a U value of 3 eV yielded the optimal band gap value, a Hubbard U value of 4 eV was chosen for its overall balance, which produces the lowest combined errors in lattice parameters and the band gap. This choice is consistent with previous DFT studies and experimental data~\cite{doi:10.1126/sciadv.aaw5685, Li2020, Li2021Apr}.

\begin{table}[h]
    \centering
    \caption{Properties of fully geometrically optimized \MBT using different U values. The bandgap ($E_{\text{Gap}}$), the percentage error in lattice parameters ($a$ and $c$) compared to the experimental values of 4.33~\AA~and 40.91~\AA, and the total energy difference ($\Delta E$) between AFM and FM magnetic phases per unit cell are listed. A $U=4~\mathrm{eV}$ is chosen due to the lowest combined errors in lattice parameters and the band gap.}
    \begin{ruledtabular}
    \begin{tabular}{c c c c c }

         U (eV) & $E_{\text{Gap}}$ (eV) & $\Delta^a$ (\%) & $\Delta^c$ (\%) & $\Delta E$ (meV) \\
         \hline
            0      & 0.122                 & $-$0.15         & $-$1.35         & 19.44            \\
            1      & 0.143                 & 0.07            & $-$1.10         & 11.28            \\
            2      & 0.147                 & 0.28            & $-$0.94         & 7.30             \\
            3      & 0.153                 & 0.42            & $-$0.75         & 4.44             \\
            4      & 0.147                 & 0.55            & $-$0.59         & 2.49             \\
            5      & 0.138                 & 0.75            & $-$0.55         & 1.64             \\
            6      & 0.127                 & 0.81            & $-$0.31         & $-$0.53          \\

    \end{tabular}
    \end{ruledtabular}
    \label{tab:U_value_pristine}
  %  \vspace{-10pt}
\end{table}

\begin{table}[h]
    \centering
    \caption{Detailed energy comparison for the 5.5\% Mn-vacancy bulk system across different Hubbard $U$ values. For the Relaxed method, both electronic and ionic degrees of freedom are fully optimized for each specific $U$. For the Fixed $U=4$, the calculations were performed using the fixed relaxed atomic coordinates obtained from the $U=4$ eV FM configuration to isolate the electronic effect to $U$. $E_{\text{FM}}$ and $E_{\text{AFM}}$ are total energies (eV), and the relative stability is defined as $\Delta E = (E_{\text{FM}} - E_{\text{AFM}})/18$ in meV per septuple layer (SL) per unit cell.}
    \begin{ruledtabular}
    \begin{tabular}{l c c c c }
         Method      & $U$ (eV) & $E_{\text{FM}}$ (eV) & $E_{\text{AFM}}$ (eV) & $\Delta E $ (meV/SL/cell) \\
         \hline
         Relaxed     & 3        & $-$586.213419        & $-$586.180332         & $-$1.84                   \\
                          & 4        & $-$581.689869        & $-$581.655863         & $-$1.89                   \\
                          & 5        & $-$577.675680        & $-$577.642964         & $-$1.82                   \\
         \hline
         Fixed $U=4$ & 3        & $-$586.177108        & $-$586.143453         & $-$1.87                   \\
                              & 4        & $-$581.689869        & $-$581.655863         & $-$1.89                   \\
                              & 5        & $-$577.673136        & $-$577.640362         & $-$1.82                   \\
    \end{tabular}
    \end{ruledtabular}
    \label{tab:U_detailed_vacancy_18}
\end{table}

\clearpage

\section{Bulk non-collinear total energies results}

\begin{table}[h]
\centering
\caption{Total energies for bulk MBT at various defect types and concentrations. The fourth column shows the results after structural optimization of the atomic positions for each magnetic configuration. The fifth column reports the energies of the same magnetic states evaluated using the ground-state relaxed atomic structure, while keeping the atomic coordinates fixed.}
\label{tab:bulk_nc_energy}
\begin{ruledtabular}
\begin{tabular}{lcccc}

System                         & Type          & Concentration (\%) & $E_{\text{relaxed}}$ (eV/SL/cell) & $E_{\text{fixed structure}}$ (eV/SL/cell) \\ 
\midrule
Pristine~$\mathrm{-~1\times1}$ & \textbf{AFM}  & 0.0                & \textbf{$-$32.821134}             & \textbf{$-$32.821134}                     \\
Pristine~$\mathrm{-~1\times1}$ & FM            & 0.0                & $-$32.819888                      & $-$32.819797                              \\
Pristine~$\mathrm{-~2\times2}$ & \textbf{AFM}  & 0.0                & \textbf{$-$32.821242}             & \textbf{$-$32.821242}                     \\
Pristine~$\mathrm{-~2\times2}$ & FM            & 0.0                & $-$32.820017                      & $-$32.819900                              \\
Pristine~$\mathrm{-~3\times3}$ & \textbf{AFM}  & 0.0                & \textbf{$-$32.821867}             & \textbf{$-$32.821867}                     \\
Pristine~$\mathrm{-~3\times3}$ & FM            & 0.0                & $-$32.820461                      & $-$32.820520                              \\
\midrule
Mn-vacancy                     & AFM           & 12.5               & $-$31.682997                      & $-$31.683518                              \\
Mn-vacancy                     & \textbf{FM}   & 12.5               & \textbf{$-$31.688415}             & \textbf{$-$31.688415}                     \\
Mn-vacancy                     & AFM           & 5.5                & $-$32.314196                      & $-$32.314215                              \\
Mn-vacancy                     & \textbf{FM}   & 5.5                & \textbf{$-$32.316104}             & \textbf{$-$32.316104}                     \\
\midrule
Intermixing                    & \textbf{AFM}  & 12.5               & \textbf{$-$32.783886}             & $-$32.783886                              \\
Intermixing                    & \textbf{AFiM} & 12.5               & $-$32.783853                      & \textbf{$-$32.783933}                     \\
Intermixing                    & FiM           & 12.5               & $-$32.782944                      & $-$32.783176                              \\
Intermixing                    & FM            & 12.5               & $-$32.782967                      & $-$32.783045                              \\
Intermixing                    & AFM           & 5.5                & $-$32.804444                      & $-$32.804315                              \\
Intermixing                    & \textbf{AFiM} & 5.5                & \textbf{$-$32.804618}             & \textbf{$-$32.804618}                     \\
Intermixing                    & FiM           & 5.5                & $-$32.803522                      & $-$32.803693                              \\
Intermixing                    & FM            & 5.5                & $-$32.803265                      & $-$32.803290                              \\
\midrule                                                  
Mn-rich                        & AFM           & 12.5               & $-$33.144876                      & $-$33.145165                              \\
Mn-rich                        & AFiM          & 12.5               & $-$33.142259                      & $-$33.141191                              \\
Mn-rich                        & FiM           & 12.5               & $-$33.144689                      & $-$33.143305                              \\
Mn-rich                        & \textbf{FM}   & 12.5               & \textbf{$-$33.147777}             & \textbf{$-$33.147777}                     \\
\midrule
Mn-rich                        & \textbf{AFM}  & 5.5                & \textbf{$-$32.971346}             & \textbf{$-$32.971346}                     \\
Mn-rich                        & AFiM          & 5.5                & $-$32.971126                      & $-$32.971228                              \\
Mn-rich                        & FiM           & 5.5                & $-$32.970102                      & $-$32.970248                              \\
Mn-rich                        & FM            & 5.5                & $-$32.970533                      & $-$32.970740                              \\
\midrule               
Bi-rich                        & \textbf{AFM}  & 12.5               & \textbf{$-$32.392826}             & \textbf{$-$32.392826}                     \\
Bi-rich                        & FM            & 12.5               & $-$32.392058                      & $-$32.392190                              \\
Bi-rich                        & \textbf{AFM}  & 5.5                & \textbf{$-$32.632257}             & \textbf{$-$32.632257}                     \\
Bi-rich                        & FM            & 5.5                & $-$32.632071                      & $-$32.631504                              \\

\end{tabular}
\end{ruledtabular}
\end{table}

\begin{figure}[h]
    \centering
    \includegraphics{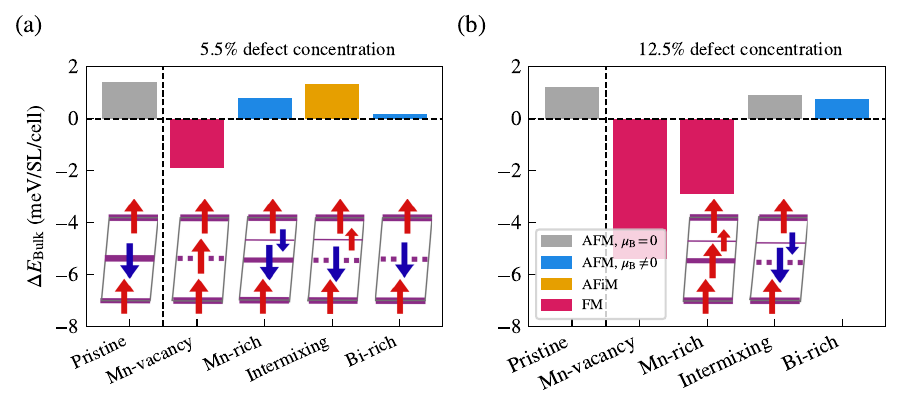} % Include the PDF figure
    \caption{Interlayer coupling for pristine and defective supercell models containing (a) 5.5\% and (b) 12.5\% defect concentrations including the energy due to relaxation of the atomic positions. Positive (negative) $\Delta E_{\mathrm{Bulk}}$ values indicate a preference for the AFM (FM) ordering. The schematic models depict the preferred magnetic ground state of the SL containing the defects for each model. Red (blue) upward (downward) arrows represent magnetic moments oriented upward (downward) within a given layer. The solid purple lines indicate Mn layers without defects, dashed lines indicate Mn layers missing one magnetic atom, and the slim purple line in the Mn-rich and intermixing models signifies the presence of an additional Mn layer.}

    \label{fig:Etotbulk}
    \begin{minipage}[t]{0.49\textwidth}
      \phantomsubcaption
      \label{fig:FMcoupling55supp}
  \end{minipage}
  \begin{minipage}[t]{0.49\textwidth}
      \phantomsubcaption
      \label{fig:FMcoupling125supp}
  \end{minipage}
\end{figure}

\clearpage

\section{Monolayer non-collinear total energies results}

\begin{table}[th]
\centering
\caption{Total energies from non-collinear calculations. The fourth column 
shows the results after structural optimization of the atomic positions for each magnetic configuration. The fifth column shows the energies of the same magnetic states evaluated using the ground-state relaxed atomic structure from either FM or FiM, while keeping the atomic coordinates fixed.}
\begin{ruledtabular}

\begin{tabular}{lcccc}

System                                       & Type         & Concentration (\%) & $E_{\text{relaxed}}$ (eV/SL/cell) & $E_{\text{fixed structure}}$ (eV/SL/cell) \\
\midrule
Intermixing-NN                               & FM           & 25                 & $-$32.374942                      & $-$32.374578                              \\
Intermixing-NN                               & \textbf{FiM} & 25                 & \textbf{$-$32.375460}             & \textbf{$-$32.375460}                     \\
Intermixing-NN                               & FM           & 11.1               & $-$32.418499                      & $-$32.418238                              \\
Intermixing-NN                               & \textbf{FiM} & 11.1               & \textbf{$-$32.419097}             & \textbf{$-$32.419097}                     \\
Intermixing-NN                               & FM           & 6.25               & $-$32.434182                      & $-$32.434128                              \\
Intermixing-NN                               & \textbf{FiM} & 6.25               & \textbf{$-$32.434703}             & \textbf{$-$32.434703}                     \\
\midrule
$\text{Intermixing-2}^{\text{nd}}\text{-NN}$ & FM           & 25                 & $-$32.398854                      & $-$32.398753                              \\
$\text{Intermixing-2}^{\text{nd}}\text{-NN}$ & \textbf{FiM} & 25                 & \textbf{$-$32.401449}             & \textbf{$-$32.401449}                     \\
$\text{Intermixing-2}^{\text{nd}}\text{-NN}$ & FM           & 11.1               & $-$32.424538                      & $-$32.424395                              \\
$\text{Intermixing-2}^{\text{nd}}\text{-NN}$ & \textbf{FiM} & 11.1               & \textbf{$-$32.424902}             & \textbf{$-$32.424902}                     \\
$\text{Intermixing-2}^{\text{nd}}\text{-NN}$ & FM           & 6.25               & $-$32.437136                      & $-$32.437051                              \\
$\text{Intermixing-2}^{\text{nd}}\text{-NN}$ & \textbf{FiM} & 6.25               & \textbf{$-$32.437374}             & \textbf{$-$32.437374}                     \\
\midrule 
$\text{Intermixing-3}^{\text{rd}}\text{-NN}$ & FM           & 25                 & $-$32.374959                      & $-$32.374612                              \\
$\text{Intermixing-3}^{\text{rd}}\text{-NN}$ & \textbf{FiM} & 25                 & \textbf{$-$32.375493}             & \textbf{$-$32.375493}                     \\
$\text{Intermixing-3}^{\text{rd}}\text{-NN}$ & \textbf{FM}  & 11.1               & \textbf{$-$32.416500}             & $-$32.416500                              \\
$\text{Intermixing-3}^{\text{rd}}\text{-NN}$ & \textbf{FiM} & 11.1               & $-$32.416397                      & \textbf{$-$32.416539}                     \\
$\text{Intermixing-3}^{\text{rd}}\text{-NN}$ & FM           & 6.25               & \textbf{$-$32.433647}             & $-$32.433647                              \\
$\text{Intermixing-3}^{\text{rd}}\text{-NN}$ & FiM          & 6.25               & $-$32.433579                      & \textbf{$-$32.433649}                     \\
\midrule
Mn-rich                                      & \textbf{FM}  & 25                 & \textbf{$-$33.101038}             & \textbf{$-$33.101038}                     \\
Mn-rich                                      & FiM          & 25                 & $-$33.077632                      & $-$33.078323                              \\
Mn-rich                                      & \textbf{FM}  & 11.1               & \textbf{$-$32.746083}             & \textbf{$-$32.746083}                     \\
Mn-rich                                      & FiM          & 11.1               & $-$32.741719                      & $-$32.742170                              \\
Mn-rich                                      & \textbf{FM}  & 6.25               & \textbf{$-$32.618691}             & \textbf{$-$32.618691}                     \\
Mn-rich                                      & FiM          & 6.25               & $-$32.616433                      & $-$32.617061                              \\

\end{tabular}
\end{ruledtabular}
\label{tab:Et_noncollinear}
\end{table}
\clearpage

\section{Relative Stability of Magnetic Configurations in Monolayer MBT}
\label{si:relative_stab_ML}

As shown in \cref*{tab:configurations_ML,tab:energy_stability_comparison}, we compared the uniform FM intralayer alignment against two configurations where local Mn spins were flipped to an antiparallel orientation relative to the rest of the layer. 
For both the pristine system and the Mn-rich supercell, the locally flipped spin configurations were higher in energy. 

% --- TABLE 2: MONOLAYER MODELS (Two-column span) ---
\begin{table}[th]
\centering
\small
\caption{Atomic structure configurations for $\text{MnBi}_2\text{Te}_4$ monolayer defect models in a $2\times2\times1$ supercell. For all monolayer models, $a = b =  8.68$~\AA, $c = 26.29$~\AA, $\alpha=\beta = 90^\circ$ and $\gamma = 120^\circ$, with internal atomic positions fully relaxed.}
\label{tab:configurations_ML}
\begin{ruledtabular}
\begin{tabular}{lcccc} 

Model                                                                                                                 & FM                                                                                                                        & FiM                                                                                                                     & AFM-3-4 & AFM-4 \\ \midrule
% --- ML Pristine ---
Pristine                                                                                                              &            
\raisebox{-.5\height}{\includegraphics[height=1.3in]{pristine-FM-2x2.pdf}}  & 
---                                                                                                                   & \raisebox{-.5\height}{\includegraphics[height=1.3in]{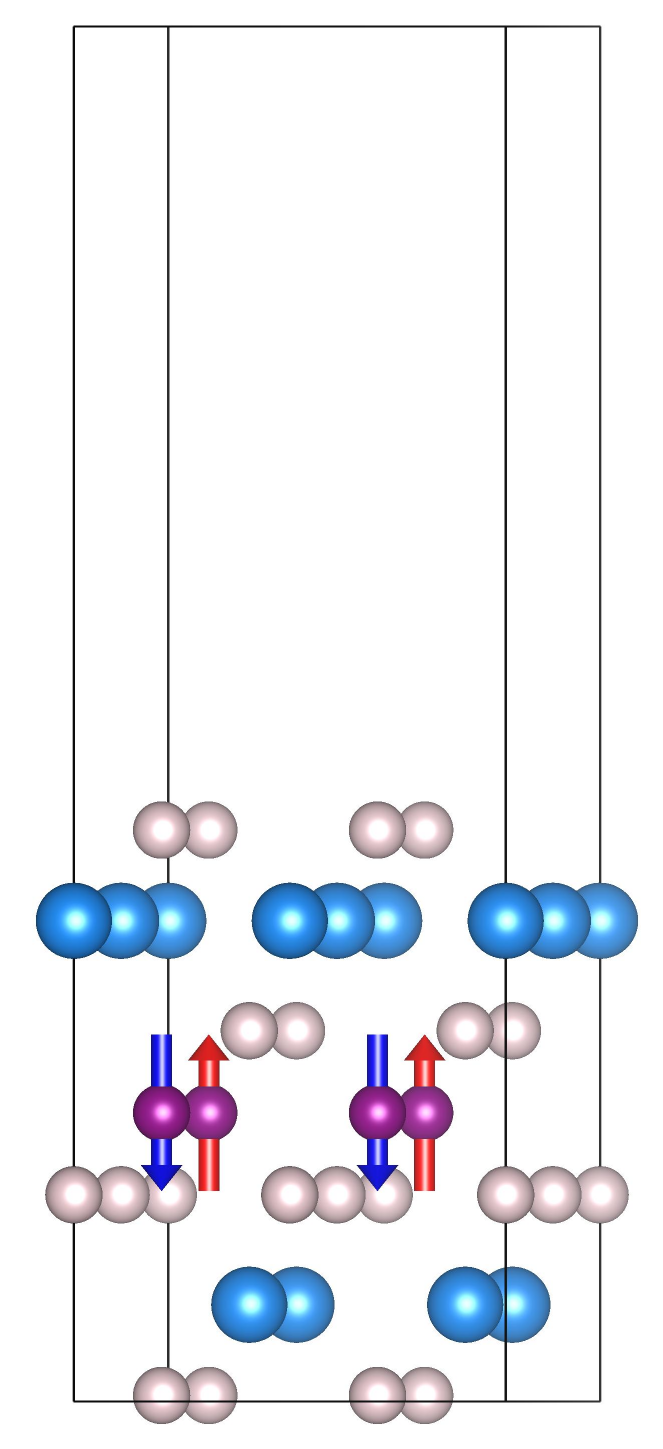}} & \raisebox{-.5\height}{\includegraphics[height=1.3in]{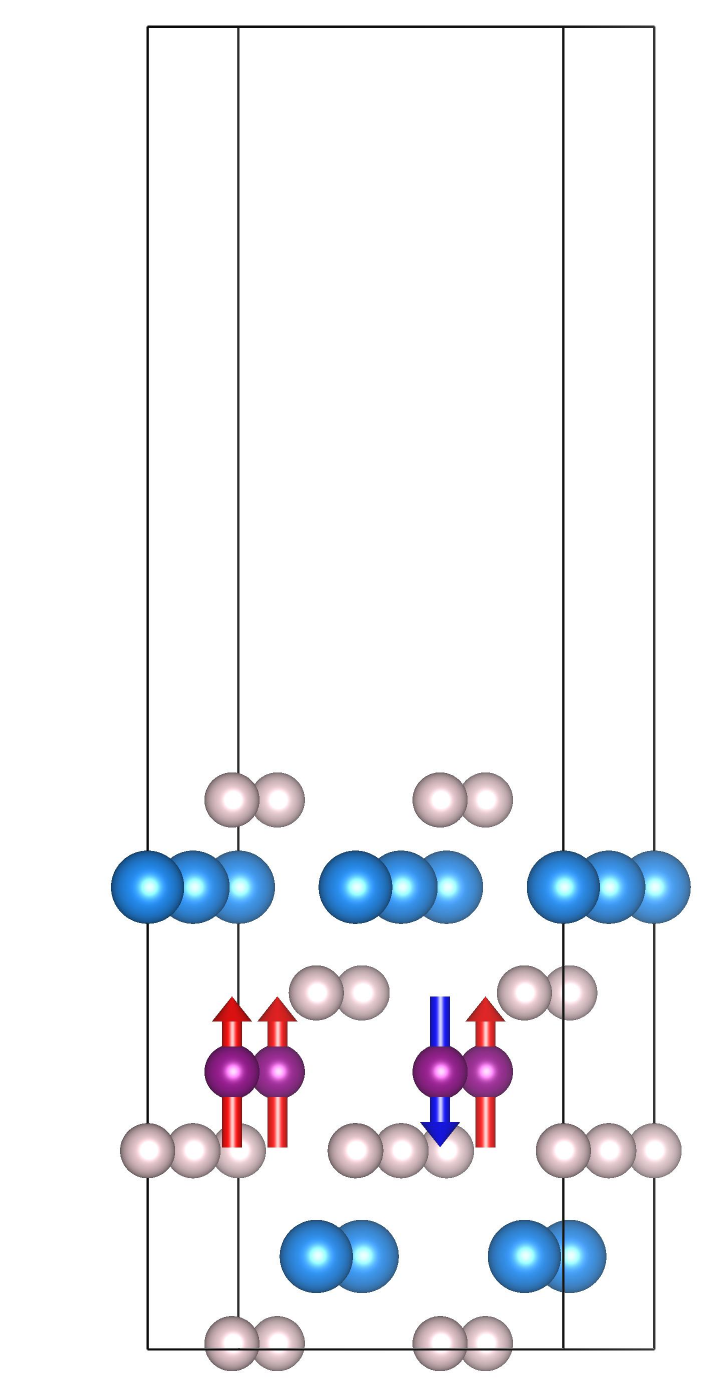}}                   \\ \midrule
% --- ML Mn-rich ---
Mn-rich                                                                                                               & 
\raisebox{-.5\height}{\includegraphics[height=1.3in]{Mn-rich-FM-1-2x2.pdf}} & 
\raisebox{-.5\height}{\includegraphics[height=1.3in]{Mn-rich-FM-2-2x2.pdf}} & 
\raisebox{-.5\height}{\includegraphics[height=1.3in]{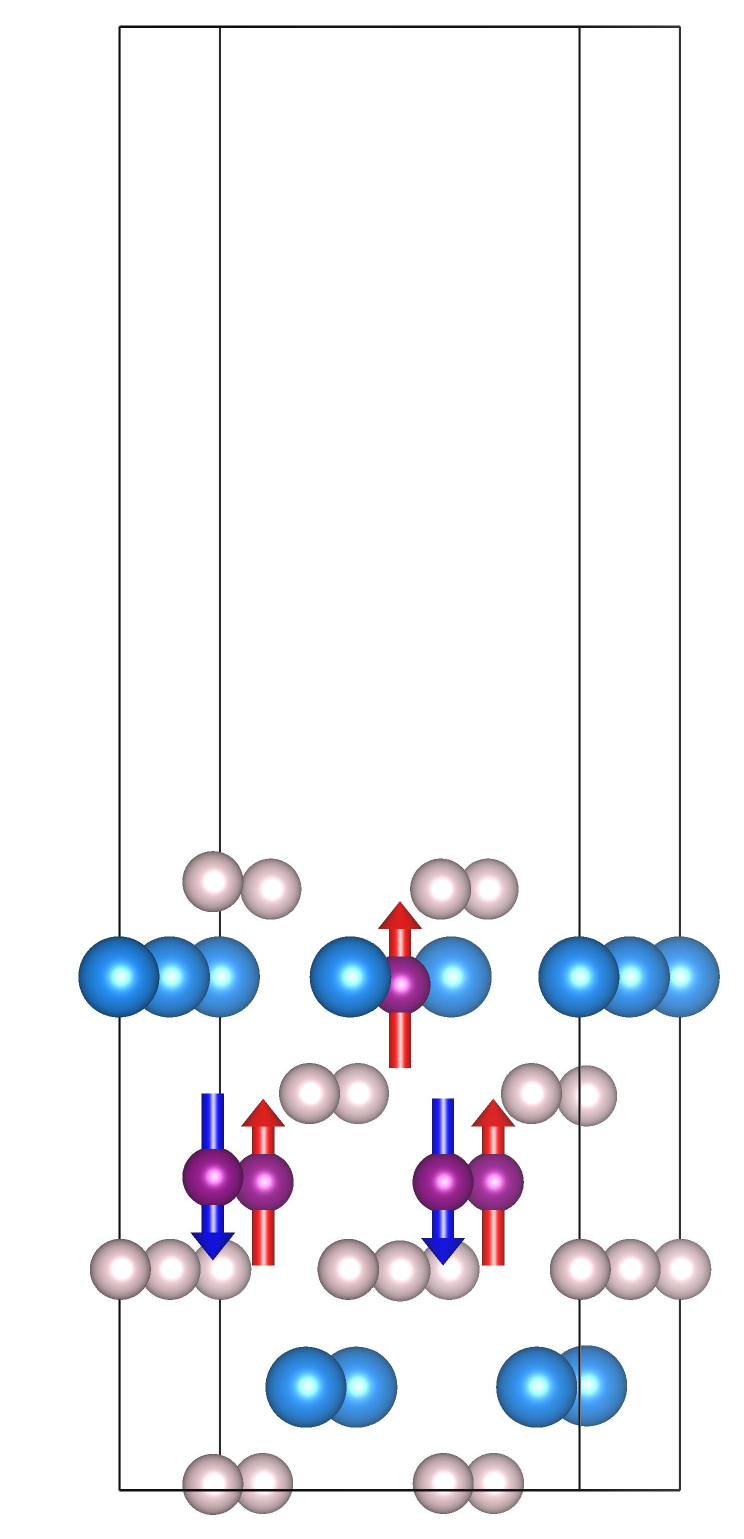}}
 & 
\raisebox{-.5\height}{\includegraphics[height=1.3in]{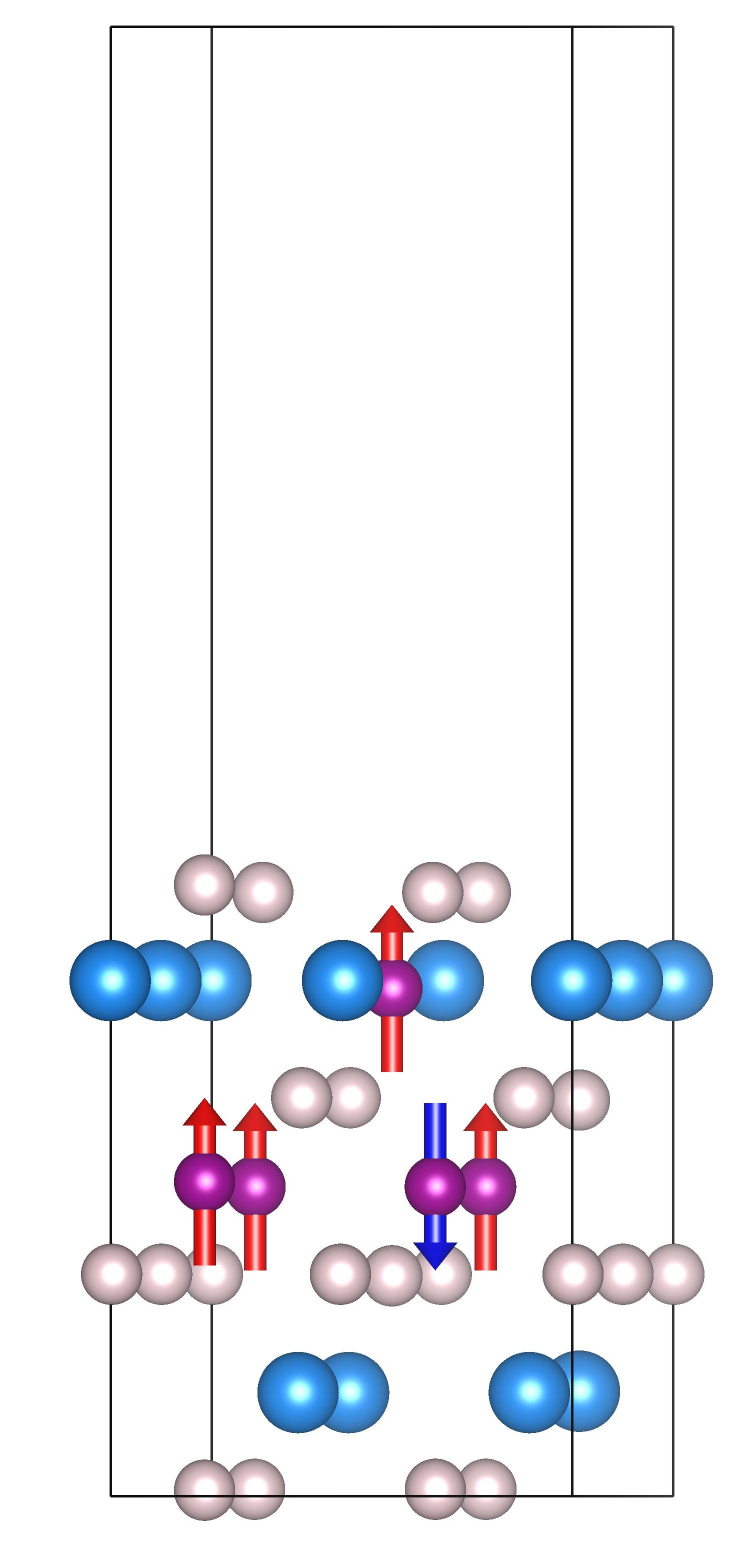}}                                                                                                                                                                                                                                                                        \\ 

\end{tabular}
\end{ruledtabular}
\end{table}

\begin{table}[h]
    \centering
    \caption{Detailed energy comparison for Pristine and Mn-rich (25\%) systems. The FM-1 configuration represents the ground state for both systems. Relative stability is calculated as $\Delta E = (E_{\text{FM-1}} - E_{\text{Config}})/4$ in meV per septuple layer per unit cell. A $\Delta E$ of 0.00 identifies the reference ground state, while negative values indicate higher energy configurations.}
    \begin{ruledtabular}
    \begin{tabular}{l l c c}

         System   & Configuration & $E_{\text{Config}}$ (eV) & $\Delta E $ (meV/SL/cell) \\
         \midrule
         Pristine & FM-1          & $-$129.810054            & 0.00                      \\
                  & AFM-3-4       & $-$129.788573            & $-$5.37                   \\
                  & AFM-4         & $-$129.793937            & $-$4.03                   \\
         \midrule
         Mn-rich  & FM-1          & $-$132.404166            & 0.00                      \\
                  & FM-2          & $-$132.310530            & $-$23.41                  \\
                  & AFM-3-4       & $-$132.296552            & $-$26.90                  \\
                  & AFM-4         & $-$132.358796            & $-$11.34                  \\
   
    \end{tabular}
            
    \end{ruledtabular}
    \label{tab:energy_stability_comparison}
\end{table}

\clearpage
\section{Monolayer projected density of states}

\begin{figure}[!ht]
    \centering
    \includegraphics[width=\textwidth]{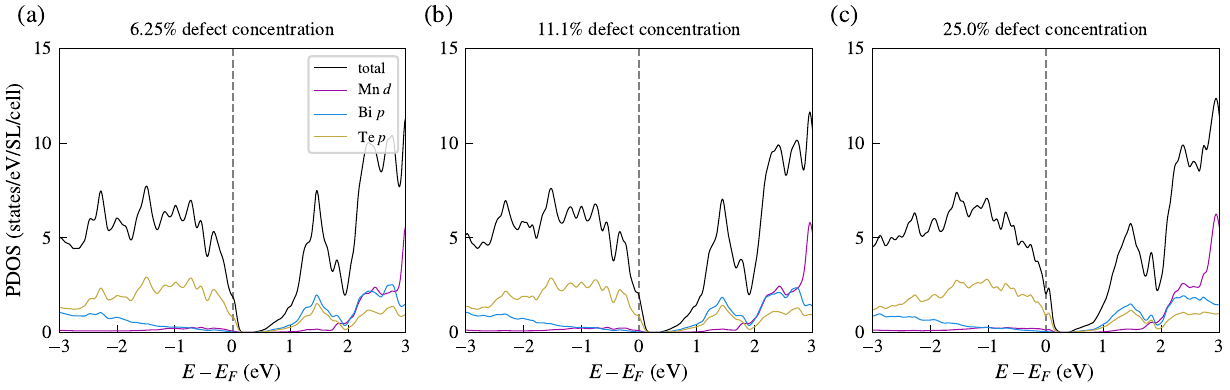}
    \caption{
        Projected density of states (PDOS) for the Mn-rich system
        in the FM ground state at (a) $6.25\%$, (b) $11.1\%$, and (c) $25.0\%$ defect concentrations. 
        The Fermi energy ($E_{\mathrm{F}}$) is set to zero. 
        The total DOS and contributions from Mn $d$, Bi $p$, and Te $p$ states are shown.
    }
    \label{fig:combined_pdos_Mn_rich}
\end{figure}

\begin{figure}[!ht]
    \centering
    \includegraphics[width=\textwidth]{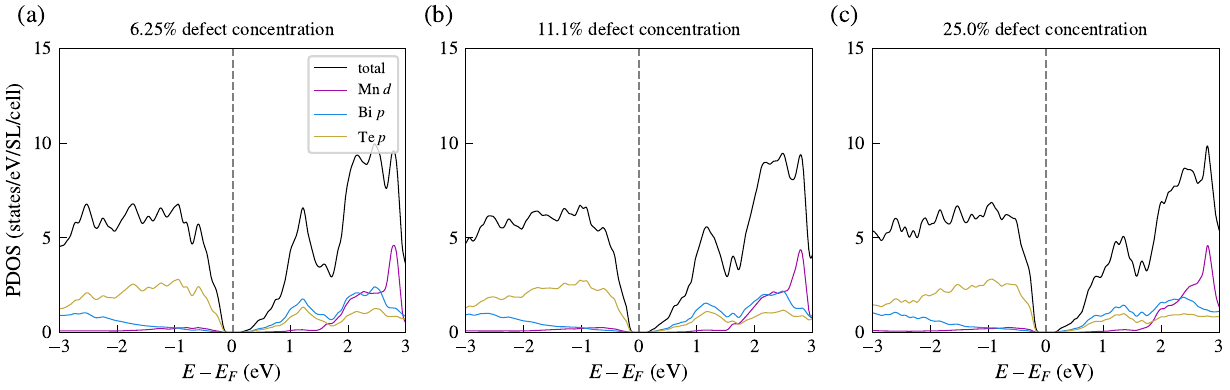}
    \caption{
        PDOS for the intermixing-NN system
        in the FiM ground state at (a) $6.25\%$, (b) $11.1\%$, and (c) $25.0\%$ defect concentrations. 
        $E_{\mathrm{F}}$ is set to zero. 
        The total DOS and contributions 
        from Mn $d$, Bi $p$, and Te $p$ states are shown.
    }
    \label{fig:combined_pdos_Intermixing}
\end{figure}

\clearpage

\section{Exchange parameters}
\subsection{Energy convergence}

\begin{figure}[ht]
    \centering
    \includegraphics[width=0.7\textwidth]{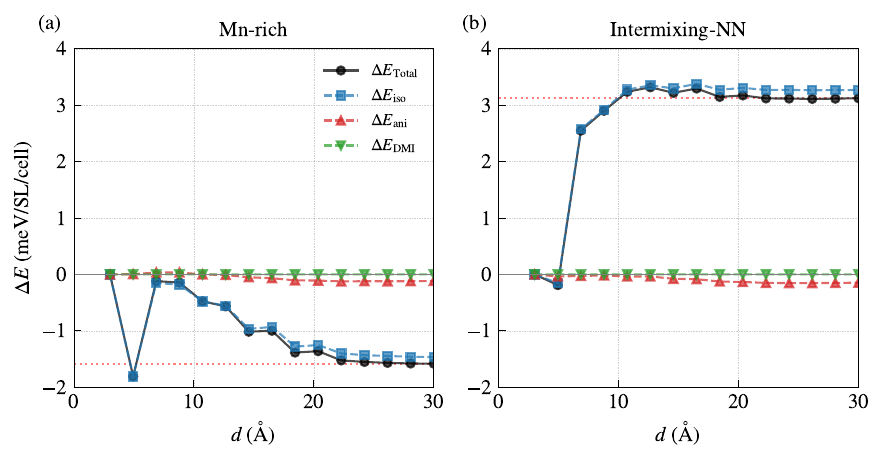}
    \caption{Convergence of the magnetic energy difference ($\Delta E$) between the FM and FiM states as a function of the interatomic cutoff distance for the (a) Mn-rich and (b) intermixing-NN at 11.1\% concentration. The total energy, incorporating isotropic exchange ($J^{\mathrm{iso}}$), symmetric anisotropic exchange ($J^{\mathrm{ani}}$), and the Dzyaloshinskii-Moriya interaction (DMI), is shown by the black solid line. The blue, red, and green dashed lines represent the individual contributions to $\Delta E$ from $J^{\mathrm{iso}}$, $J^{\mathrm{ani}}$, and DMI, respectively. The horizontal red dotted line denotes the converged total energy value. The close alignment between the isotropic energy and the total energy reveals that $J^{\mathrm{iso}}$ is the dominant contributor to the $\Delta E$.}
    \label{fig:Econv_Mn-richandintermix}
\end{figure}

\subsection{Mn-rich 11.1\%}
\begin{figure}[ht]
    \centering
    \includegraphics[width=0.5\textwidth]{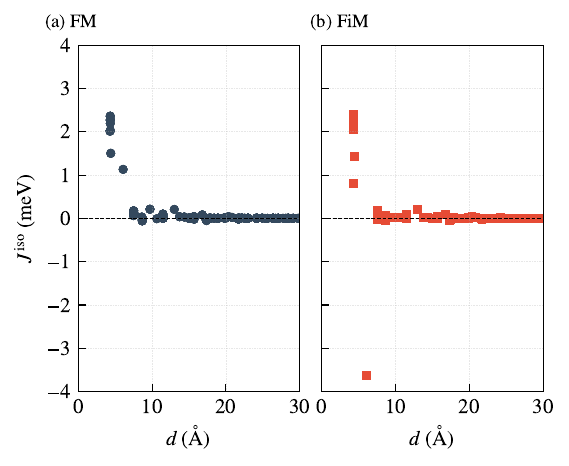}
    \caption{Isotropic exchange ($J^{\mathrm{iso}}$) as a function of the distance for (a) FM and (b) FiM configurations for the Mn-rich system at 11.1\% defect concentration.}
    \label{fig:Mn-richJiso}
\end{figure}
\begin{figure}[ht]
    \centering
    \includegraphics[width=\textwidth]{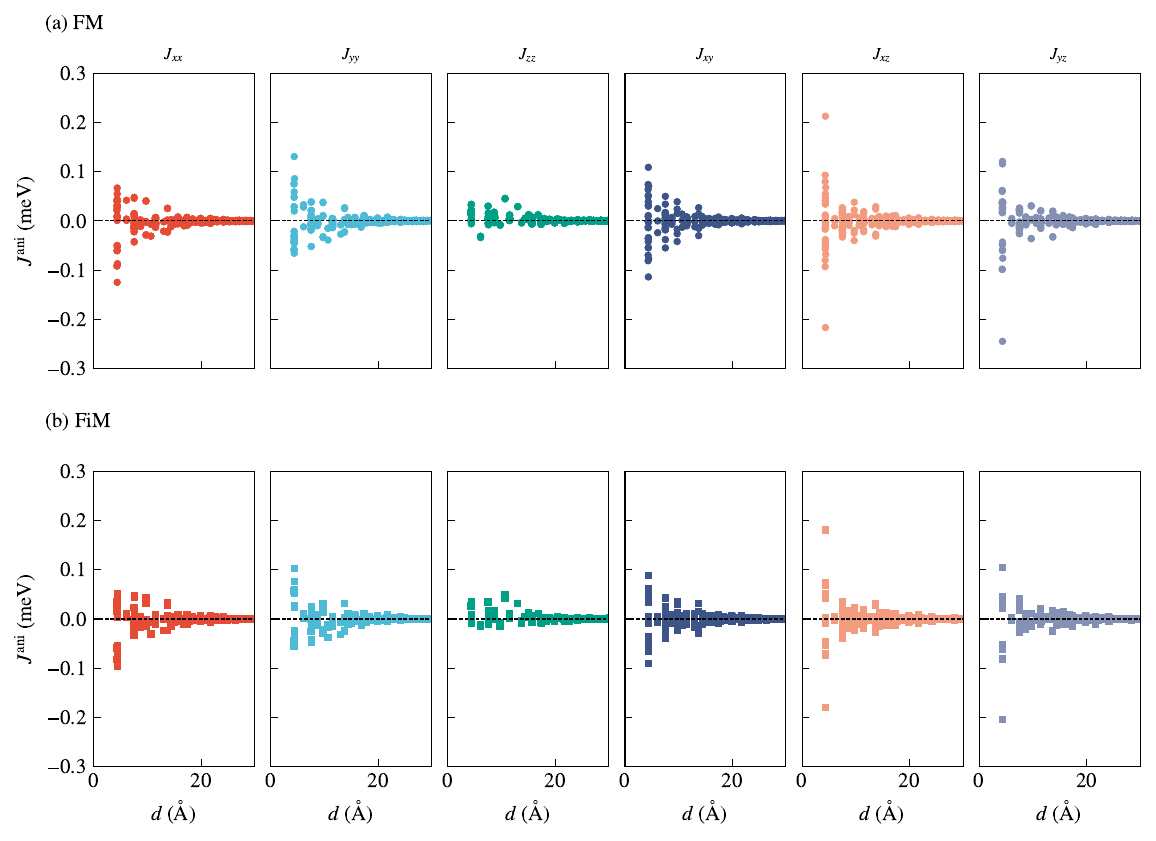}
    \caption{Symmetric anisotropic exchange ($J^{\mathrm{ani}}$) tensor components as a function of the distance for (a) FM and (b) FiM configurations for the Mn-rich system at 11.1\% defect concentration.}
    \label{fig:Mn-richJani}
\end{figure}

\begin{figure}[ht]
    \centering
    \includegraphics[width=\textwidth]{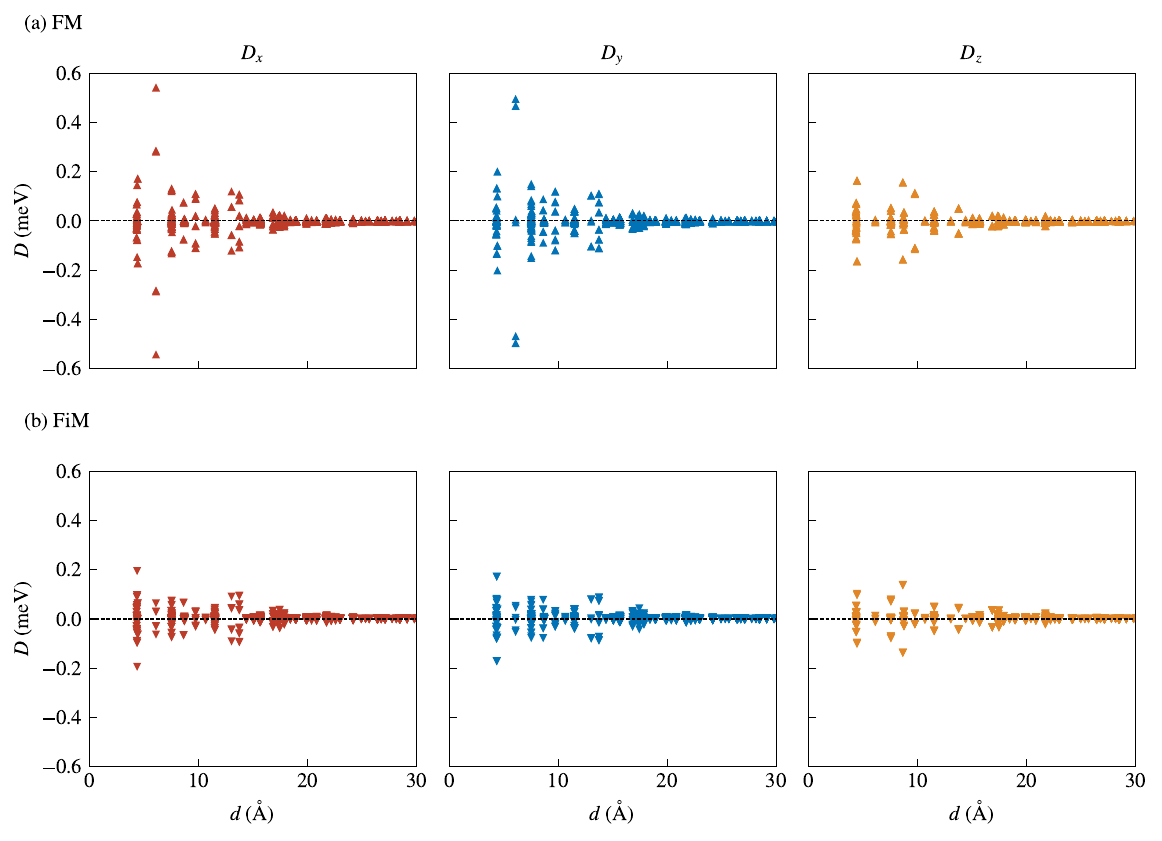}
    \caption{Dzyaloshinskii-Moriya interaction (DMI) vector components as a function of interatomic distance for (a) the FM state and (b) the FiM state for the Mn-rich system at 11.1\% defect concentration. The plots illustrate the $D_x$, $D_y$, and $D_z$ components, with values ranging between $-$0.6 and 0.6 meV.}
    \label{fig:Mn-richDMI}
\end{figure}

\clearpage

\subsection{Intermixing 11.1\%}

\begin{figure}[ht]
    \centering
    \includegraphics[width=0.5\textwidth]{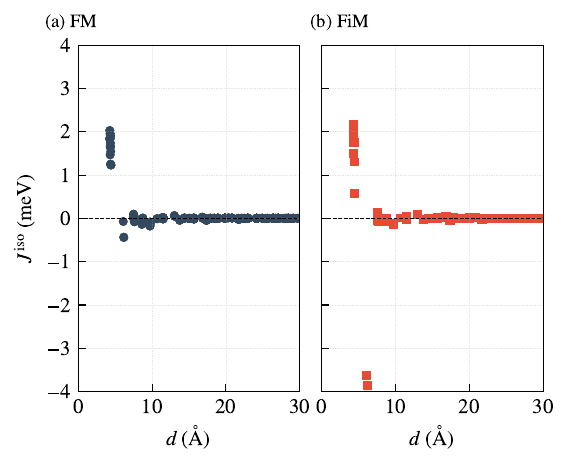}
    \caption{Isotropic exchange ($J^{\mathrm{iso}}$) as a function of the distance for (a) FM and (b) FiM configurations for the intermixing system at 11.1\% defect concentration.}
    \label{fig:intermixingJiso}
\end{figure}

\begin{figure}[ht]
    \centering
    \includegraphics[width=\textwidth]{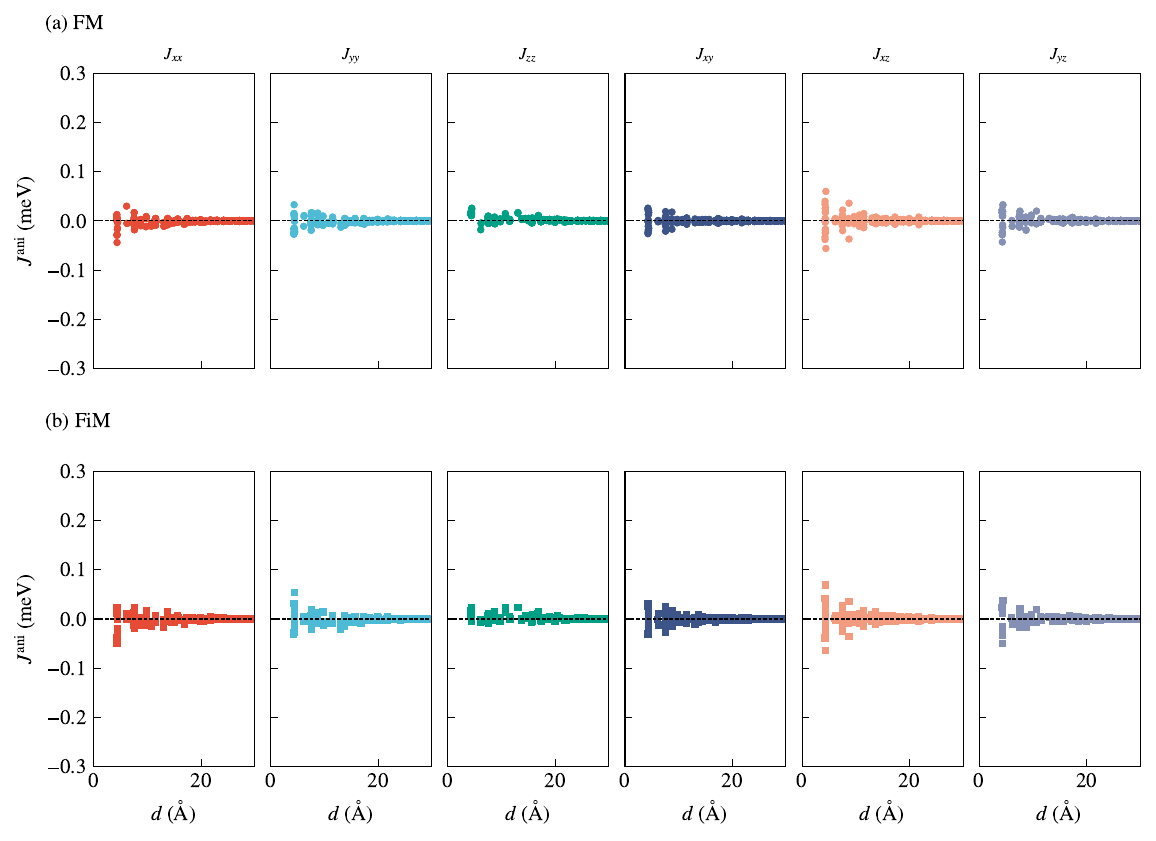}
    \caption{Symmetric anisotropic exchange ($J^{\mathrm{ani}}$) tensor components as a function of the distance for (a) FM and (b) FiM configurations for the intermixing-NN system at 11.1\% defect concentration.}
    \label{fig:intermixingJani}
\end{figure}

\begin{figure}[ht]
    \centering
    \includegraphics[width=\textwidth]{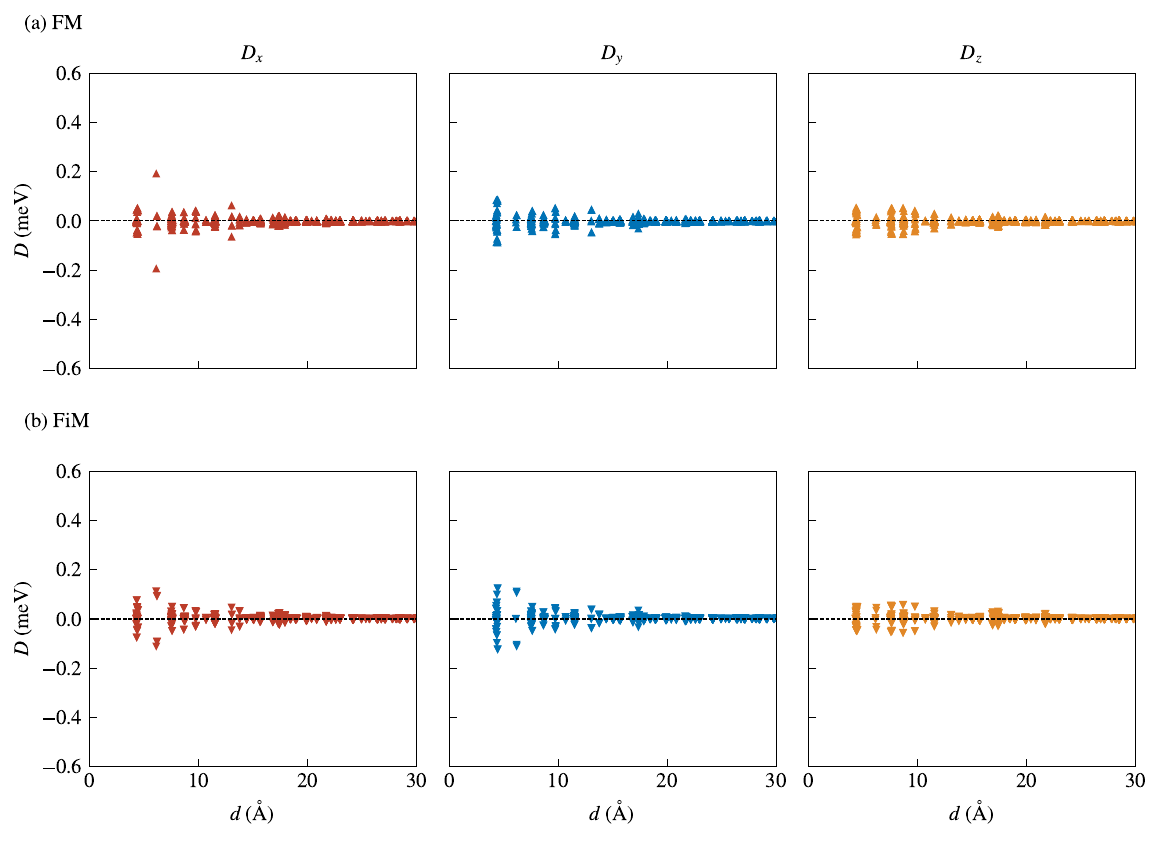}
    \caption{Dzyaloshinskii-Moriya interaction (DMI) vector components as a function of interatomic distance for (a) the FM state and (b) the FiM state for the intermixing-NN system at 11.1\% defect concentration. The plots illustrate the $D_x$, $D_y$, and $D_z$ components, with values ranging between $-$0.2 and 0.2 meV.}
    \label{fig:intermixingDMI}
\end{figure}

\clearpage

\bibliography{ref.bib}